\documentclass[10pt,a4paper]{article}
\usepackage[utf8]{inputenc}
\usepackage[english]{babel}
\usepackage{indentfirst}
\usepackage{textcomp}
\usepackage{authblk}
\usepackage[numbers, sort&compress]{natbib}
\usepackage[autostyle]{csquotes}
\usepackage{mathtools}

\usepackage[left=2.5cm,right=2.5cm,top=2cm,bottom=2cm]{geometry}

\usepackage{graphicx}
\usepackage{amsmath}
\usepackage{amssymb}
\usepackage[version=4]{mhchem}
\usepackage{siunitx}
\usepackage{longtable,tabularx}
\usepackage{soul}

\usepackage{comment}
\usepackage[x11names]{xcolor}
\usepackage{hyperref}
\hypersetup{
    colorlinks=true,
    linkcolor=red,
    citecolor=blue,
    }
\usepackage[capitalize]{cleveref}

\usepackage{tikz}
\usepackage{pgfplots}
\pgfplotsset{compat=1.9}
\usetikzlibrary{calc}
\usetikzlibrary{arrows.meta}
\usetikzlibrary{patterns,hobby}
\usetikzlibrary{angles, quotes}
\usepackage[percent]{overpic}
\usepackage{subcaption}

\pgfmathsetmacro{\rot}{15}
\pgfmathsetmacro{\px}{0}
\pgfmathsetmacro{\py}{0}

\newtheorem{theorem}{Theorem}

\title{\bf Physically consistent formulation for the bound vortex sheet strength in the Wagner model}

\author[1]{George Lucas S. Torres\footnote{Ph.D. Candidate, Department of Mechanical Engineering, george.torres@usp.br (Corresponding Author).}}
\author[2]{Ashok Gopalarathnam\footnote{Professor, Department of Mechanical and Aerospace Engineering, agopalar@ncsu.edu.}}
\author[1]{Flávio D. Marques\footnote{Associate Professor, Department of Mechanical Engineering, fmarques@sc.usp.br.}}
\affil[1]{\small São Carlos School of Engineering, University of São Paulo, São Carlos, SP, 13566-590, Brazil}
\affil[2]{\small North Carolina State University, Raleigh, NC 27695, USA}

\date{\small \underline{Accepted manuscript}}

\begin{document}

\maketitle

\begin{center}
\small
\textit{This article has been accepted for publication in \textit{Physical Review Fluids}. DOI: https://doi.org/10.1103/hcd2-znt5}
\end{center}

\begin{abstract}
Unsteady thin-airfoil theory (UTAT) coupled with discrete-vortex methods has been widely employed in reduced-order aerodynamic modeling. Due to the non-uniqueness of potential-flow solutions, the Kutta condition is imposed to determine the circulation around the airfoil. Although the unsteady Kutta condition is commonly associated with the zero-loading condition at the trailing edge, its implications for the continuity of the vortex-sheet strength remain comparatively underexplored. In particular, the classical series expansion employed for the bound vorticity in unsteady thin-airfoil theory is not uniformly convergent at the trailing edge, leading to mathematical inconsistencies in the vortex-sheet and pressure distributions. In this context, the present work seeks to advance the mathematical framework of unsteady thin-airfoil theory through a physically consistent formulation of the bound vortex-sheet strength for the Wagner problem. A recurrence relation is derived for the Wagner coefficients, allowing the construction of a uniformly convergent series expansion for the bound vorticity. The proposed formulation ensures continuity between the bound and wake vortex sheets while simultaneously recovering zero pressure loading at the trailing edge, thereby providing a mathematically consistent representation of the unsteady Kutta condition. To investigate the implications of the modified framework, a discrete-vortex method based on UTAT is developed and compared with the classical formulation. The results demonstrate that the proposed approach eliminates spurious oscillatory behavior near the trailing edge and significantly improves the regularity of the computed vorticity and pressure distributions. Although both formulations appear to approach Wagner’s classical lift response at large times, the classical representation remains inconsistent during the transient stage due to the discontinuity of the vortex-sheet strength. The proposed methodology establishes a more robust theoretical foundation for reduced-order unsteady aerodynamic models and provides a consistent basis for future extensions involving arbitrary airfoil kinematics.

\end{abstract}

\section{Introduction}

Theoretical and numerical formulations of unsteady aerodynamic phenomena seen in nature are constantly evolving for academic and industrial purposes. Recent applications find a place in flapping-wing aerodynamics of insects and birds \citep{chin2016flapping}, environmental monitoring and homeland security through micro-air vehicles (MAVs) \citep{shyy2010recent}, bio-inspired flight, such as perching and hovering maneuvers \citep{narsipur2023discrete}, onshore and offshore vertical-axis wind turbines (VAWTs) \citep{hand2020review}, to name a few. Computational fluid dynamics (CFD) simulations and experimental investigations have been conducted to extract the primary characteristics and physics behind these unsteady phenomena \citep{deparday2019modeling,izquierdo2021experimental,deparday2022experimental,miotto2022analysis,daliri2023experimental,camacho2024real}. In addition, Laplace equation solutions for inviscid incompressible flow problems are widely used in the study of lifting surfaces, and linear superpositions of two-dimensional potential flow solutions such as vortices and sources are a traditional modeling approach \citep{ramesh2014discrete,liu2017discrete,taha2020high,narsipur2020variation,sureshbabu2021theoretical,saini2021leading,martinez2022modulation,ramanathan2023prediction}.

Early developed models paved the way for a better understanding of unsteady aerodynamics. \citet{wagner1925entstehung} presented circulatory responses due to a sudden acceleration of a flat plate. \citet{theodorsen1935general} treated the potential flow problem of a harmonically oscillating airfoil through the complex analysis of conformal mappings. \citet{garrick1936propulsion} modeled the propulsion arising from the leading-edge suction on oscillating wings based on Theodorsen's formulas and the method outlined by \citet{karman1935general}. \citet{kussner1936zusammenfassender} obtained solutions for a sharp-edged gust, and \citet{sears1938systematic} solved the problem of sinusoidal transverse gusts. \citet{jones1938operational} developed a two-state approximation of the Wagner function based on operational methods, which can be related to the Theodorsen function through a Fourier transform \citep{garrick1938some}. \citet{vepa1976use} applied Padé approximants to obtain a finite-state representation of unsteady aerodynamic loads in the frequency domain of completely arbitrary time-dependent motions. Moreover, \citet{isaacs1945airfoil} extended Theodorsen's work to compute the aerodynamic loads of an airfoil in a variable stream magnitude but a constant angle of attack. In contrast, \citet{greenberg1947airfoil} obtained forces and moments on a two-dimensional airfoil that undergoes sinusoidal motions in a pulsating stream.

Contemporary unsteady aerodynamic models are based on these noteworthy classical linear theories developed over the past century. \citet{motta2015influence} explored the influence of airfoil thickness on aerodynamic loads for harmonically pitching airfoils and proposed a modification of the Theodorsen model to take into account the effects of airfoil thickness. \citet{strangfeld2016airfoil} conducted theoretical and experimental investigations of an airfoil in high-amplitude harmonic oscillations of the free stream at a constant angle of attack. \citet{otomo2021unsteady} adapted the linear theory of Theodorsen and the unsteady thin-airfoil theory to accurately predict the lift of unsteady airfoils under massively separated flows and non-sinusoidal kinematics. \citet{gennaretti2022kutta} presented the extension of the Kutta–Joukowski theorem to unsteady linear aerodynamics based on the Theodorsen frequency response function. \citet{tiomkin2022unsteady} analytically examined lift responses of aeroelastic extensions to Theodorsen, Wagner, Sears, and Küssner functions for a membrane airfoil subjected to small-amplitude chord motions and transverse gusts.

Drawing attention back to classical models, further methods solve the aerodynamic problem directly within the airfoil domain rather than applying conformal mapping. \citet{schwarz1940berechnung} applied the inversion formula of \citet{sohngen1939losungen} to solve the integral equation of the no-penetration condition for the bound and wake vortex sheet strengths in harmonic oscillations. For arbitrary motions, \citet{katz1978behavior} enhanced thin-airfoil theory \cite{glauert1926elements} by implementing the discrete-vortex time-stepping method to compose the unsteady thin-airfoil theory (UTAT). More recently, \citet{ramesh2011augmentation} introduced the concept of the leading edge suction parameter (LESP) into a large-amplitude maneuver version of UTAT, titled the LESP-modulated discrete-vortex method (LDVM), allowing an intermittent leading edge vortex (LEV) formation based on the suction force in the airfoil's leading edge. \citet{ramesh2020leading} investigated the principle of matched asymptotic expansions (MAE) to derive closed-form expressions for the velocity at the leading edge and the location of the stagnation point in airfoils undergoing arbitrary motions. \citet{narsipur2023low} presented a reduced-order method to account for dynamic-stall effects based on the LDVM and the unsteady trailing-edge separation method (UTM) \citep{narsipur2019low}.

\subsection{The bound vorticity}

Primarily, the classical works focused on obtaining aerodynamic loads and the circulation history, with somewhat little attention paid to the bound vortex sheet. However, modern techniques, such as viscous-inviscid interaction methods \cite{veldman2009simple,ramos2014strong,ramos2016three,paturle2022dynamic}, rely on an accurate representation of the bound vorticity distribution. In this approach, the boundary layer is computed using a viscous model. In contrast, the velocity over the airfoil surface, computed from potential flow theories, is frequently employed as the outer solution. In simulating finite wings, strip theory provides solutions for large aspect ratios and is considered one of the simplest approaches known in the literature \cite{phlips1981unsteady,izraelevitz2017state,boutet2018unsteady}. Recently, \citet{bird2021unsteady} presented an unsteady lifting-line model derived using the principle of MAE. This solution incorporates the inner solution from Theodorsen's theory, based on the classical two-dimensional bound vorticity formulation.

Accordingly, theoretical and computational efforts have been made to predict the bound vorticity distribution in two-dimensional flow simulations. In pioneering work on unsteady airfoil theory, \citet{sears1938systematic} introduced the coefficients $Q_n$ related to the wake-induced velocity to derive the bound vorticity distribution for harmonically oscillating airfoils. Sears calculated the pressure distribution on the airfoil using a recurrence formula for $Q_n$. In this context, the computation of the bound vorticity distribution requires suitable summation terms. However, obtaining the coefficients is not straightforward, as the recurrence formula does not converge for large $n$ or small reduced frequencies \cite{epps2018vortex}. To overcome this problem, \citet{epps2018vortex} developed a numerical method to calculate the vortex sheet strength in the aerodynamic problems of Wagner, Theodorsen, Küssner, and Sears. Recently, \citet{ramesh2025correct} presented closed-form expressions for modeling the shed vorticity through the vorticity distribution derived from the Wagner solution. However, the classical series expansion for the bound vorticity employed in these models does not converge uniformly at the trailing edge.

In the classical thin-airfoil theory, \citet{glauert1926elements} anticipates the bound vorticity distribution as a sum of a singular part and a Fourier sine series to ensure zero vorticity at the trailing edge. However, as argued by \citet{rienstra1992note}, this is only true if the series tends to zero continuously near the trailing edge and not only pointwise. By interpreting Glauert's approach in the context of generalized functions \citep{jones1982theory}, Rienstra presented uniformly convergent series expansions for the bound vorticity distribution. Despite the mathematical validity of Glauert's approach in steady-flow regimes, its application to unsteady aerodynamic problems presents meaningful challenges related to the Kutta condition.

\subsection{The Kutta condition}
\label{sec:unsteady_Kutta_condition}

Solutions for potential flows are not unique for a given set of boundary conditions, and the well-known Kutta condition \citep{crighton1985kutta} is applied to provide suitable airfoil circulation. This additional condition has been postulated in several ways in recent decades, summarized as the necessary condition imposed on the flow to leave the trailing edge smoothly \citep{xia2017unsteady}. In this context, the validity of results obtained from reduced-order models hinges on the accuracy with which the underlying physical considerations and governing equations are formulated and implemented \citep{katz2001low}.

\citet{roesler2018discretization} established discretization criteria for the unsteady vortex-lattice method (UVLM) by analyzing the influence of time-step, bound vortex spacing, and wake vortex spacing to reproduce Wagner's results. Using the Wagner analytical solution as a benchmark, they demonstrated that inadequate discretization results in an incorrect representation of the Kutta condition and, consequently, erroneous force predictions. When developing and validating a first-order solution for UTAT, \citet{ramesh2020leading} demonstrated that applying the method of MAE can provide useful closed-form expressions for crucial aerodynamic features in the leading-edge region. Although an excellent agreement between Euler CFD and Theodorsen simulations is observed for the airfoil's overall pressure distribution, a significant discrepancy arises when attempting to accurately predict the pressure difference near the trailing edge employing the classical UTAT formulation.

Robust approaches focus on the desingularization of suitable functions when simulating two-dimensional vortex sheets to avoid numerical instabilities, as in \citet{krasny1986desingularization}. Within this technique, \citet{jones2003separated} removed the inverse square-root and logarithmic singularities in the velocity potential of a moving plate. \citet{alben2010regularizing} investigated the convergence rate of such desingularized functions and proposed two numerical methods that decrease the error relative to existing methods. However, bound vorticity formulations for canonical unsteady models were not presented, as their model considers two vortex sheets emanating from both airfoil edges at a high effective angle of attack.

Although the zero-loading condition at the trailing edge seems to be a well-established requirement for satisfying the Kutta condition in the literature, the requirement for continuous vortex sheet strength across the trailing edge has been less thoroughly investigated. In the context of thin-airfoil theory, \citet{eldredge2019mathematical} presents theoretical and mathematical foundations for the \textit{unsteady} Kutta condition in two-dimensional inviscid incompressible flows, significantly based on the works of \citet{jones2003separated} and \citet{alben2010regularizing}. Two consequential statements declare that \textit{i)} a continuous pressure is expected in the vicinity of an edge, which eliminates the possibility of pressure difference on either side of the body as the edge is approached and that \textit{ii)} the strength of the bound vortex sheet is equal to the strength of the free vortex sheet at an edge of zero angle, and their common value is generally nonzero.

Despite recent progress in theoretically and computationally predicting bound vorticity with reduced-order models, a significant gap persists in the literature regarding a rigorous mathematical framework that accurately represents the physics of unsteady flows. Despite the extensive efforts of \citet{jones2003separated,alben2010regularizing,eldredge2019mathematical,ramesh2025correct}, the existing bound-vorticity formulations based on Glauert's thin-airfoil theory remain in need of improvement. In essence, a more robust theoretical foundation for the bound vortex sheet strength is required to address unsteady aerodynamic problems, particularly in light of the Kutta condition, i.e., to ensure the validity of statements \textit{i)} and \textit{ii)} mentioned earlier.

\subsection{Proposed formulation}

The present work seeks to advance unsteady airfoil theory from a mathematical standpoint, particularly by improving the treatment of vortex-sheet continuity across the trailing edge. Given the role of classical aerodynamic models as benchmarks for validating more sophisticated approaches, the Wagner problem, which describes the airfoil response to a sudden acceleration, provides a simple and suitable case study. Such an investigation sheds light on possible improvements toward more general kinematics.

This work introduces a physically consistent bound-vorticity formulation for the Wagner model that properly accounts for the unsteady Kutta condition. A recurrence formula is obtained for the Wagner coefficients $R_n$, which can be readily replaced within the classical formula of bound vorticity. A dubious series emerges, and convergence analyses demonstrate that it is the source of the mathematical inconsistency in light of the unsteady Kutta condition. A straightforward test is applied to the modified series expansion to validate its uniform convergence, yielding a mathematically consistent formulation for the bound vorticity. 

Hence, a simple discrete-vortex method (DVM) based on UTAT is developed to investigate the impacts of the modified formulation on aerodynamic loads. Inspired by the proposed analytical framework, a consistent discrete-vortex formulation is derived for computing the bound vorticity. The new methodology allows for faster convergence than the classical approach since only a few terms need to be summed to predict the bound-vorticity distribution accurately.

\section{Theory}
\label{sec:classical_theory}

This Section provides the theoretical basis for some reduced-order models, including a discussion of the applicability of the unsteady Kutta condition. \cref{sec:UTAT} introduces the unsteady thin-airfoil theory (UTAT) and presents the aerodynamic loads in terms of Fourier coefficients $A_n$. The classical Wagner problem is briefly introduced in \cref{sec:Wagner}, where its Fourier coefficients are denominated $R_n$. \cref{sec:kutta_condition} explores the theoretical and mathematical foundations of the unsteady Kutta condition and its implementation in reduced-order models. The discrete-vortex version of the Wagner model is developed in \cref{sec:DVM}, where the implementation of the unsteady Kutta condition follows the previous analytical framework.

\subsection{Unsteady thin-airfoil theory}
\label{sec:UTAT}

Unsteady thin-airfoil theory provides a robust framework for simulating airfoils undergoing arbitrary low-amplitude motions and the natural roll-up of wake vortices (see \citet{katz2001low} for details). While modern approaches for UTAT have been developed to accommodate large-amplitude maneuvers and leading-edge vortex shedding (see \citet{ramesh2014discrete}), the classical version is employed here to facilitate a consistent comparison with the small-disturbance model explored later, namely the Wagner model.

\begin{figure}[!htbp]
	\centering
	\begin{tikzpicture}[>=stealth]
		\begin{axis}[
			axis equal,
			width=0.6\linewidth,
			height=0.25\linewidth,
			xmin=0, xmax=1.75,
			axis line style={draw=none},
			ticks=none,
			clip=false,
			anchor=origin,
			]

			\draw[->,rotate=-\rot] (axis cs: 0,0) -- (axis cs: 1.2,0)
			node [pos=1, anchor=north] {$x$};
			;
			\draw[->,rotate=-\rot] (axis cs: 0,0) -- (axis cs: 0,0.5)
			node [pos=1, anchor=south] {$z$}
            node[pos=0, anchor=south east] {$b$}
			;
			\addplot [black!30, fill=gray,fill opacity=0.2, rotate = -\rot] table 			{table/naca0012.txt};
			
			\draw [fill=black, rotate = -\rot] (axis cs: 0.25,0) circle [radius=0.01];
			
			\draw[->, rotate = -\rot, thick, Red2] (axis cs: 0.25,0) +(-30:10) arc (-30:-330:10)
			node [pos=0.3,anchor=north east] {$\dot{\alpha}$} 
			;
			
			\draw[->,rotate = -\rot, Red2, thick] (axis cs: 0.25,0) -- (axis cs: {0.25-0.5*cos(\rot)},{-0.5*sin(\rot)}) 
			node [pos=1,anchor=east] {$U$}
			;
			\draw[->,rotate = -\rot, Red2, thick] (axis cs: 0.25,0) -- (axis cs: {0.25-0.2*sin(\rot)},{0.2*cos(\rot)}) 
			node [pos=1,anchor=south] {$\dot{h}$}
			;
			
			\draw[->, rotate = -\rot, Tan4] (axis cs: {0.25-0.35*cos(\rot)},{-0.35*sin(\rot)}) arc [radius=35,start angle=-180+\rot,end angle=-180]
			 node [pos=0.5,anchor=east] {$\alpha$}
			;
			
			\draw[densely dotted, thick, rotate = -\rot] (axis cs: 0,0) -- (axis cs: -0.25,0);
			
			\draw[densely dotted, thick, rotate = -\rot] (axis cs: 0.25,0) -- (axis cs: 0.25,0.4);
			
			\draw[<->, rotate = -\rot, Tan4] (axis cs: 0,0.35) -- (axis cs: 0.25,0.35)
			node [pos=0.5,anchor=south,rotate=-\rot] {$ac$}
			;

            \draw[->, rotate = -\rot] (axis cs: 0.75,0) -- (axis cs: 0.75,0.15)
            node [pos=1, anchor=south, font=\scriptsize] {$v_n(x,t)$}
            ;
            
			\draw [SpringGreen4, thick,rotate=-\rot] (axis cs: 1,0) to [ curve through ={(axis cs: 1.1,0.075) . . (axis cs: 1.1,0.25) .. (axis cs: 1.4,0.55)  }] (axis cs: 1.75,0.8);

		\end{axis}
	\end{tikzpicture}
	\caption{Airfoil kinematics.}
	\label{fig:airfoil_kinematics}
\end{figure}
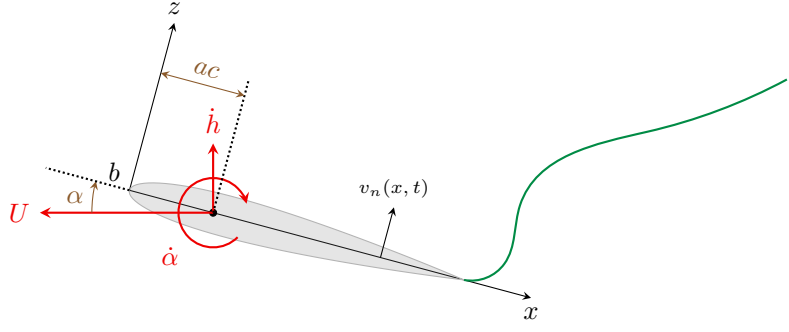

The airfoil kinematics is shown schematically in \cref{fig:airfoil_kinematics}, where $c$ is the airfoil chord, $U$ is the horizontal velocity, $\alpha$ is the angle of attack, $\dot{h}$ is the upward velocity, and $x=ac$, $0\leq a\leq 1$, is the pivot location. The vortex sheet lies on the $x$-axis when assuming small-disturbance approximations, and the normal induced velocity by a vorticity distribution $\gamma(x,t)$ is given by
\begin{equation} \label{eq:dphi_dz}
    \dfrac{\partial \phi}{\partial z}(x,z = 0,t) = - \dfrac{1}{2\pi} \int_{\mathcal{C}} \dfrac{\gamma(\xi,t)}{x-\xi} d\xi,
\end{equation}
where $\phi(x,z,t)$ is the velocity potential and $\mathcal{C}$ is the integration path. For a fully developed wake, the no-penetration condition reads
\begin{equation} \label{eq:no-penetration}
    -\dfrac{1}{2\pi} \int_0^c \dfrac{\gamma_b(\xi,t)}{x-\xi} d\xi - \dfrac{1}{2\pi} \int_c^\infty \dfrac{\gamma_w(\xi,t)}{x-\xi} d\xi = v_n(x,t),
\end{equation}
where $\gamma_b$ is the airfoil (bound) vorticity, $\gamma_w$ is the wake vorticity, and $v_n$ is the normal velocity of the airfoil:
\begin{equation} \label{eq:normal_velocity}
	v_n(x,t) = \dot{h} - U\alpha + \dot{\alpha}(ac-x).
\end{equation}
The contribution of the camber line is neglected in this analysis for the sake of brevity. However, a method for obtaining closed-form expressions for general airfoil surfaces and their integration into the thin-airfoil theory is presented by \citet{torres2024nonlinear}.

As in \citet{glauert1926elements}, the vorticity distribution over the airfoil is taken as a trigonometric expansion,
\begin{equation} \label{eq:glauert_vorticity}
	\gamma_b(\theta,\tau) = 2 U \left[A_0 (\tau)\cot{(\theta/2)} + \sum_{n=1}^\infty A_n(\tau) \sin(n\theta) \right],
\end{equation}
where $A_0$, $A_1$,$\ldots$, $A_n$ are the Fourier series coefficients, $\tau=(2U/c)t$ is the non-dimensional time, and $\theta$ and $x$ are related as
\begin{equation} \label{eq:Glauert_transform}
    x = \dfrac{c}{2} (1-\cos \theta).
\end{equation} 
In this classical expansion, the bound vorticity at the trailing edge is
\begin{equation} \label{eq:bound_vorticity_TE}
    \gamma_{b_{\text{TE}}}(\tau) := \gamma_b(\pi,\tau) \equiv 0.
\end{equation}

The coefficients $A_n$ in \eqref{eq:glauert_vorticity} are obtained through Glauert's principal-value integrals \citep{legua2017cauchy},
\begin{align}
	\int_0^\pi \dfrac{\cos{(n\vartheta)}}{\cos\vartheta - \cos\theta} d\vartheta &= \pi\dfrac{\sin{(n\theta)}}{\sin{\theta}}, \quad n = 0,1,2,\ldots , \label{eq:glauert1} \\
	\int_0^\pi \dfrac{\sin{(n\vartheta)}\sin{\vartheta}}{\cos\vartheta - \cos\theta} d\vartheta &= -\pi\cos{(n\theta)}, \quad n = 1,2,3,\ldots, \label{eq:glauert2}
\end{align}
as:
\begin{align}
	A_0(\tau) &=  -\dfrac{v_{1/2}}{U} + \dfrac{1}{\pi U} \int_0^\pi \dfrac{\partial \phi_w}{\partial z} \, d\theta, \label{eq:Fourier_0} \\
    A_1(\tau) &=  \alpha'  -\dfrac{2}{\pi U} \int_0^\pi \dfrac{\partial \phi_w}{\partial z} \cos\theta \, d\theta, \label{eq:Fourier_1} \\
	A_{n\geq 2}(\tau) &= - \dfrac{2}{\pi U} \int_0^\pi  \dfrac{\partial \phi_w}{\partial z} \cos {(n\theta)}  \, d\theta, \label{eq:Fourier_n}
\end{align}
where $(\cdot)' = d(\cdot)/d\tau$, $v_{1/2}(\tau) = 2U\left[h'/c - \alpha/2 + \alpha'(a-1/2)\right]$, and $\partial \phi_w / \partial z$ is the $z$-component of the velocity induced by the wake. The bound circulation is evaluated by the integration of the bound vorticity \eqref{eq:glauert_vorticity} over the chord:
\begin{align} \label{eq:airfoil_circulation}
	\Gamma_b(\tau) = \pi c U \left( A_0 + \dfrac{A_1}{2} \right).
\end{align}

\subsubsection{Aerodynamic loads}

To compute the aerodynamic loads, Euler's equation is integrated along a streamline --- also called \textit{unsteady} Bernoulli equation \citep{katz2001low} --- to obtain the pressure distribution, that is:
\begin{equation} \label{eq:unsteady_bernoulli}
    \dfrac{\partial \phi}{\partial t} + \dfrac{V^2}{2} + \dfrac{p}{\rho} = \mathcal{F},
\end{equation}
where $\mathcal{F}$ is a function of time along a streamline, $\rho$ is the fluid density, and $V=|\nabla \phi|$.

The tangential velocity induced by a vortex distribution on its surface is given by
\begin{equation} \label{eq:dphi_dx}
    \dfrac{\partial \phi_{u,l}}{\partial x}(x,t) = \lim_{z\to 0^\pm} \dfrac{\partial \phi}{\partial x}(x,z,t) = \pm \dfrac{1}{2}\gamma (x,t),
\end{equation}
where $\phi_u$ and $\phi_l$ are the velocity potential on the upper and lower surfaces of the vortex sheet, respectively. Then, applying \cref{eq:unsteady_bernoulli} on the upper and lower surfaces of the airfoil leads to the pressure difference distribution
\begin{equation}
    \dfrac{\Delta p}{\rho}(x,t) = \dfrac{v_u^2 - v_l^2}{2} + \dfrac{\partial }{\partial t}(\phi_u - \phi_l),
\end{equation}
where $v_u$ and $v_l$ are the tangential velocity on the upper and lower surfaces of the airfoil, respectively: 
\begin{equation}
    v_{u,l}(x,t) = U \pm \dfrac{\gamma_b (x,t)}{2}.
\end{equation}

Denoting $\phi_{u,l}(0,t) \equiv \phi_{LE}(t)$, the velocity potential can be written in terms of the vortex sheet strength \eqref{eq:dphi_dx} as
\begin{equation} \label{eq:velocity_potential_final}
	\phi_{u,l}(x,t) = \phi_{LE}(t) \pm \dfrac{1}{2} \int_0^x \gamma_b(x_0,t) \, dx_0,
\end{equation}
and the pressure difference distribution has the form
\begin{equation} \label{eq:pressure_katz1}
	\dfrac{\Delta p}{\rho}(x,t) =  U\gamma_b(x,t) + \dfrac{\partial }{\partial t} \int_0^{x} \gamma_b (x_0,t) \, dx_0, \quad x \in [0,c].
\end{equation}

In contrast to the bound vorticity, a free wake does not admit a pressure jump across it:
\begin{equation} \label{eq:pressure_free_wake}
	\dfrac{\Delta p_w(x,t)}{\rho} =  U\gamma_w(x,t) + \Dot{\Gamma}_b(t) + \dfrac{\partial }{\partial t} \int_c^x \gamma_w (x_0,t) \, dx_0 \equiv 0, \quad x \geq c.
\end{equation}
When developing the methodology to obtain the bound vorticity, \citet[Chapter X]{sears1938systematic} introduced the ``effective vorticity distribution" to compute the airfoil's pressure distribution:
\begin{equation} \label{eq:eff_vorticity}
	\Delta p(x,t) = \rho U \gamma_{\text{eff}}(x,t); \quad \gamma_{\text{eff}}(x,t) =  \gamma_b(x,t) + \dfrac{1}{U} \dfrac{\partial }{\partial t} \int_0^{x} \gamma_b (x_0,t) \,  dx_0,
\end{equation}
which is essentially an alternative representation of \cref{eq:pressure_katz1}. \cref{eq:pressure_free_wake} has the solution \cite{peters1995finite}
\begin{equation} \label{eq:gammaw_solution}
    \gamma_w(x,t) = - \dfrac{\Dot{\Gamma}_b(\hat{t})}{U}, \quad \hat{t} \equiv t + \dfrac{c-x}{U},
\end{equation}
which is also obtained by applying the Leibniz integral rule in Kelvin's circulation theorem,
\begin{equation} \label{eq:Kelvin_theorem}
    \dfrac{D \Gamma}{Dt}=0,
\end{equation}
where $\Gamma$ is the total circulation. \cref{eq:Kelvin_theorem} is suitable for discrete-vortex methods. Moreover, the rate at which the vorticity is shed into the wake is given by
\begin{equation} \label{eq:rate_vorticity}
    \gamma_{w_{\text{TE}}}(t) := \gamma_w(c,t) = - \dfrac{\dot{\Gamma}_b(t)}{U},
\end{equation}
or by the dimensionless form
\begin{equation} \label{eq:rate_vorticity2}
    \gamma_{w_{\text{TE}}}(\tau) = - \dfrac{\Gamma'_b}{c/2} .
\end{equation}

The pressure difference coefficient can be written in terms of the variable $\theta$ through \eqref{eq:Glauert_transform} as
\begin{equation} \label{eq:pressure_katz2}
	\Delta C_p(\theta,\tau) := \dfrac{\Delta p}{\frac{1}{2}\rho U^2} = \dfrac{2}{U}\left[ \gamma_b(\theta,\tau) +  \dfrac{\partial }{\partial \tau} \int_0^\theta  \gamma_b(\theta_0,\tau) \sin\theta_0 \,d\theta_0 \right],
\end{equation}
where the integral reads
\begin{align} \label{eq:pressure_katz2_added}
	\dfrac{1}{U}  \dfrac{\partial}{\partial \tau} \int_0^\theta \gamma_b(\theta_0,\tau) \sin\theta_0 \,d\theta_0 =& \, 2A'_0 (\theta + \sin\theta) + A'_1 \left[ \theta - \dfrac{\sin{(2\theta)}}{2} \right] \nonumber \\
	& + \sum_{n=2}^\infty A'_n \left[ \dfrac{\sin{\left[(n-1)\theta \right]}}{n-1}  - \dfrac{\sin{\left[(n+1)\theta \right]}}{n+1}\right],
\end{align}
which leads to the lift coefficient
\begin{equation} \label{eq:normal_force}
    C_l(\tau) = 2 \pi \left( A_0 + \dfrac{A_1}{2}\right) + \pi \left( 3A'_0 + A'_1 + \dfrac{A'_2}{2} \right).
\end{equation}

Within this classical framework, the pressure difference at the trailing edge is given by
\begin{equation}
    \Delta C_{p_{\text{TE}}}(\tau) := \Delta C_p(\pi,\tau) = 4\pi \left(A_0' + \dfrac{A_1'}{2} \right) ,
\end{equation}
or through \cref{eq:airfoil_circulation} by
\begin{equation} \label{eq:pressure_TE}
    \Delta C_{p_{\text{TE}}}(\tau) = \dfrac{4\Gamma'_b}{cU}.
\end{equation}
As $\Gamma'_b(\tau) \neq 0$ in unsteady flows, the wake vortex sheet strength at the trailing edge, $\gamma_{w_{\text{TE}}}$ in \cref{eq:rate_vorticity2}, is usually nonzero. In contrast, the classical formulation for bound vorticity, \cref{eq:glauert_vorticity}, yields a constant zero value at the trailing edge, thereby imposing a mathematical discontinuity in the vorticity, i.e, $\gamma_{b_{\text{TE}}}(\tau) \neq \gamma_{w_{\text{TE}}}(\tau)$. Moreover, the same discontinuity is observed in the pressure distribution across the trailing edge. Since a free wake does not admit a pressure jump across it, the trailing-edge loading should be identically zero, in contrast to \cref{eq:pressure_TE}.

These discontinuities are the source of the deviation from the unsteady Kutta condition when series expansions are employed. This topic will be covered in \cref{sec:kutta_condition}. In the next Section, the Wagner model is explored.

\subsection{The Wagner model}
\label{sec:Wagner}

\citet{wagner1925entstehung} provided a function that describes the transient lift on an airfoil that experiences sudden acceleration from rest. The model is valuable for understanding unsteady aerodynamic phenomena such as those encountered during aircraft maneuvers and gusts. Theoretical and computational  predictions of the vortex sheet strength in the seminal unsteady aerodynamic problem of Wagner were realized in the work presented by \citet{epps2018vortex}, where the Fourier coefficients $A_n$ have the form
\begin{align}
         A_0(\tau) &= \alpha_0 \left[1 - R_0(\tau)\right], \label{eq:A0_Wagner} \\
         A_{n\geq 1}(\tau) &= 2(-1)^n \alpha_0 R_n(\tau), \label{eq:An_Wagner}
\end{align}
in which $\alpha_0$ is a constant angle of attack. The coefficients $R_n$ are given by
\begin{equation} \label{eq:Rn}
    R_n(\tau) = \dfrac{1}{2\pi} \int_{-\infty}^\infty Q_n(k)S(k)e^{ik\tau}dk,
\end{equation}
where $Q_n(k)$ are the Sears coefficients \cite{sears1938systematic},
\begin{equation} \label{Qn_general}
    Q_n(k) = \int_0^\infty e^{-ik\cosh\vartheta} e^{-n\vartheta} \, d\vartheta,
\end{equation}
$S(k)$ is the Sears function,
\begin{equation}
    S(k) = \dfrac{1/ik}{K_0(ik) + K_1(ik)},
\end{equation}
and $K_n$ is the modified Bessel function of the second kind:
\begin{equation} \label{eq:Bessel}
    K_n(z) = \int_0^\infty e^{-z\cosh\vartheta} \cosh(n\vartheta) \, d\vartheta.
\end{equation}

The first two coefficients $R_n$ are readily related to well-known aerodynamic functions:
\begin{align}
    R_0(\tau) &= 1 - \Phi(\tau), \label{eq:R0} \\
    R_1(\tau) &= \Phi(\tau) - \Psi(\tau), \label{eq:R1}
\end{align}
where $\Phi$ and $\Psi$ are the Wagner and Küssner functions, respectively:
\begin{align}
    \Phi(\tau) &= \dfrac{1}{2\pi} \int_{-\infty}^\infty \dfrac{C(k)}{ik} e^{ik\tau} dk, \\
    \Psi(\tau) &= \dfrac{1}{2\pi} \int_{-\infty}^\infty \dfrac{S(k)e^{ik}}{ik} e^{ik\tau} dk,
\end{align} 
and $C(k)$ is the Theodorsen function:
\begin{equation} \label{eq:Theodorsen_function}
    C(k) = \dfrac{K_1(ik)}{K_0(ik) + K_1(ik)}.
\end{equation}
Moreover, the first two coefficients $Q_n$ read
\begin{align}
    Q_0(k) &= K_0(ik), \label{eq:Q0} \\
    Q_1(k) &= K_1(ik) - \dfrac{1}{ik}e^{-ik}. \label{eq:Q1}
\end{align}

Wagner and Küssner functions are fundamental theoretical and mathematical tools in unsteady aerodynamics, and their accurate computation can be challenging, requiring the development of suitable numerical methods and algebraic approximations (see \citet{dawson2022improved}). In this work, these functions are described as inverse Laplace transforms,
\begin{align}
    \Phi(\tau) &= \mathcal{L}^{-1}\left\{ \dfrac{C(-is)}{s} \right\}, \\
    \Psi(\tau) &= \mathcal{L}^{-1}\left\{ \dfrac{e^{-s}}{s}S(-is) \right\},
\end{align} 
and the term ``exact solution" employed in the following Sections refers to the solution obtained through numerical inversion of the Laplace transform using Euler's algorithm \cite{abate2006unified}.

The pressure difference and lift coefficients for a flat plate can be computed as
\begin{align}
    \Delta C_p(\theta,\tau) &= 4 \alpha_0 \Phi(\tau) \cot{(\theta/2)}, \label{eq:Wagner_Cp} \\
    C_l(\tau) &=  2\pi \alpha_0 \Phi(\tau), \label{eq:Wagner_Cl}
\end{align}
and the bound vorticity distribution is obtained by replacing the coefficients $A_n$ in \cref{eq:glauert_vorticity} with the expressions from \cref{eq:A0_Wagner,eq:An_Wagner}:
\begin{equation} \label{eq:Wagner_bound_vorticity}
    \gamma_b(\theta,\tau) = 2U \alpha_0 \left\{ \left[1 - R_0(\tau)  \right] \cot{(\theta/2)}  + 2\sum_{n=1}^\infty (-1)^n R_n(\tau)\sin(n\theta)\right\},
\end{equation}
where the bound circulation reads
\begin{equation} \label{eq:wagner_gamma}
    \Gamma_b(\tau) = \pi c U \alpha_0 \Psi(\tau).
\end{equation}

The wake vorticity at the trailing edge is easily obtained through \cref{eq:rate_vorticity2,eq:wagner_gamma}: 
\begin{equation} \label{eq:wake_vorticity_Wagner}
    \gamma_{w_{\text{TE}}}(\tau) = -2\pi U \alpha_0 \Psi'(\tau).
\end{equation}
Thus, the initial conditions for the bound circulation and the wake vorticity at the trailing edge are
\begin{align} \label{eq:initial_value_bound_circ_wake_vorticity}
    \Gamma_b(0) = 0, \quad \gamma_{w_{\text{TE}}}(0) \to -\infty.
\end{align}
Furthermore, the trailing-edge loading, $\Delta C_p(\pi,\tau)$ from \cref{eq:Wagner_Cp}, is zero. Indeed, in his seminal work, \citet{wagner1925entstehung} took into account the shed-vorticity rate, \cref{eq:rate_vorticity2}, in his model.

Nevertheless, mathematically, the vortex sheet strength is not yet continuous across the trailing edge as $\gamma_{b_{\text{TE}}}(\tau) \equiv 0 \neq \gamma_{w_{\text{TE}}}(\tau)$, which violates the unsteady Kutta condition. This suggests that a more appropriate series expansion should replace the classical formulation.

\subsection{Unsteady Kutta condition}
\label{sec:kutta_condition}
\pgfplotsset{scaled x ticks=false}
\pgfplotsset{scaled y ticks=false}

The traditional enforcement of the Kutta condition is known as the removal of the inverse-square-root velocity singularity at the trailing edge \citep{crighton1985kutta} and has been applied in several unsteady aerodynamic models. \citet{eldredge2019mathematical} explores the theoretical and mathematical formulations of the unsteady two-dimensional inviscid incompressible flows. In the context of the thin-airfoil theory, we quote Eldredge \cite[p.164]{eldredge2019mathematical}:
\begin{displayquote}
\textit{That the pressure is finite follows from the same requirement on the velocity. On the other hand, the requirement that the pressure be single-valued seems to have little relevance to the condition as originally stated. However, it eliminates solutions that satisfy the Kutta condition but for which there exists a discontinuity in the pressure in the fluid. [...] It should be also expected that the pressure to be continuous in the vicinity of the edge, which eliminates the possibility of pressure difference on either side of the body as the edge is approached. In other words, there is zero loading at the edge.}
\end{displayquote}
Furthermore, an important outcome of the unsteady Kutta condition is \cite[p.166]{eldredge2019mathematical}
\begin{displayquote}
\textit{For the case of an edge with zero angle, both the free and bound vortex sheet that meet at this edge represent jumps in tangential fluid velocity. It should be expected that, when the flow is regularized at the edge, the velocity is not only finite but also continuous there. As already mentioned, this velocity is not zero. In other words, the strength of the bound vortex sheet is equal to the strength of the free vortex sheet at an edge of zero angle, and their common value is generally nonzero.}
\end{displayquote}
It is straightforward to note that for steady aerodynamics, both requirements stated above are achieved through the classical mathematical approach detailed in \cref{sec:UTAT}. In this context, ensuring the Kutta condition in unsteady thin-airfoil theory relies on the accurate representation of functions using trigonometric series expansions. 

\begin{figure}[!htpb]
    \centering
    \newsavebox{\boxFourierexample}
    \savebox{\boxFourierexample}
    {
    \begin{tikzpicture}[>=stealth]
    	\begin{axis}[
    	width=0.4\linewidth,
    	height=0.2\linewidth,
    	xmin=0.8, xmax=1,
        ymin=2, ymax=4,
        ytick = {2,pi,4},
        xtick = {0.8,1},
        yticklabels={2,$\pi$,4},
        yticklabel pos=right,
        legend style={font=\scriptsize},
        clip = true,
    	]

        \addplot[Red2, samples = 2, domain=0:1] {pi*x};

        \addplot[DodgerBlue4, densely dashed, thick] table [x = theta,y=S1,col sep=comma] {table/Fourier_example.txt};

        \addplot[black, dashdotted, very thick] table [x = theta,y=S2,col sep=comma] {table/Fourier_example.txt};
    	
    	\end{axis}
    \end{tikzpicture}
    }
    \begin{tikzpicture}[>=stealth]
    	\begin{axis}[
    	width=0.8\linewidth,
    	height=0.45\linewidth,
        xlabel={$x / \pi$},
    	xmin=0, xmax=2,
        ymin=-4, ymax=4,
        ytick = {-4,0,4},
        xtick = {0,1,2},
        legend style={font=\scriptsize},
        clip = true,
        xmajorgrids=true,
        ymajorgrids=true,
    	]

        \addplot[Red2,samples = 2, domain=0:1] {pi*x};
        \addlegendentry{$f(x)=x$ on $[0,\pi]$}

        \addplot[DodgerBlue4, densely dashed, thick] table [x = theta,y=S1,col sep=comma] {table/Fourier_example.txt};
        \addlegendentry{$f_{\text{odd}}(x)$ (\cref{eq:Fourier_example1})}

        \addplot[black, dashdotted, very thick] table [x = theta,y=S2,col sep=comma] {table/Fourier_example.txt};
        \addlegendentry{$f_{\text{even}}(x)$ (\cref{eq:Fourier_example2})}

        \draw (axis cs: 0.5,-2) node{\usebox{\boxFourierexample}};
    	
    	\end{axis}
    \end{tikzpicture}
    \caption{Odd and even Fourier series expansions of the function $f(x)=x$.}
    \label{fig:Fourier_example}
\end{figure}
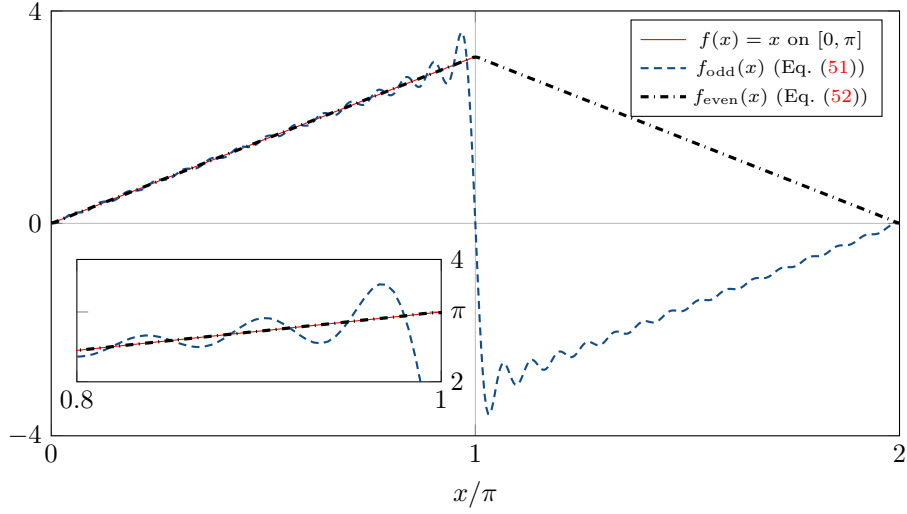

To illustrate that, one takes the Fourier series
\begin{equation} \label{eq:Fourier_example1}
    f_{\text{odd}}(x) = -2\sum_{n=1}^\infty \dfrac{(-1)^n}{n} \sin(nx),
\end{equation}
which is zero at $x=\pi$, but behaves like $\sim x$ for $x \to \pi$. \cref{eq:Fourier_example1} represents the series expansion of the sawtooth wave $f(x) = x$ at the interval $[-\pi,\pi]$ and converges to $[f(x-)+f(x+)]/2 = 0$ at the discontinuity points. Now, consider the series expansion
\begin{equation} \label{eq:Fourier_example2}
    f_{\text{even}}(x) = \dfrac{\pi}{2} -\dfrac{4}{\pi}\sum_{n=1}^\infty \dfrac{\cos\left[(2n-1)x\right]}{(2n-1)^2},
\end{equation}
which represents the Fourier series expansion of a triangular wave. \cref{fig:Fourier_example} compares the sawtooth and triangular waves taking $n=30$ terms on the interval $[0,2\pi]$. Although both series can represent the function $f(x)=x$ in the half-open interval $[0,\pi)$, $f_{\text{even}}(x)$ converges quickly and does not exhibit oscillatory behavior at the discontinuity points, known as the Gibbs phenomenon \citep{thompson1992fourier}.

To advance in obtaining valid series for the proposed problem, the concepts of \textit{pointwise convergence} and \textit{uniform convergence} for a series of functions are explored. Evidently, the series $f_{\text{odd}}(x)$ does not converge pointwise to $f(x)=x$ on $\mathcal{D} := \{ x \in [0,\pi] \}$, since $f_{\text{odd}}(\pi) = 0 \neq f(\pi)$. On the other hand, the series $f_{\text{even}}(x)$ is likely to represent $f(x)$ on $\mathcal{D}$ since $f_{\text{even}}(\pi) = \pi = f(\pi)$. However, a more powerful statement than pointwise convergence is needed, since the rate of convergence can vary significantly with $x$ in $\mathcal{D}$, as shown in \cref{fig:Fourier_example}. Hence, the definition of uniform convergence of infinite series is presented \cite[p.189]{abbott2015understanding}:
\begin{theorem}[Cauchy Criterion for Uniform Convergence]
    A series $\sum_{n=1}^\infty f_n(x)$ converges uniformly on $\mathcal{I} \subseteq \mathbb{R}$ if and only if for every $\varepsilon > 0$ there exists an $N \in \mathbb{N}$ such that
    \begin{equation*}
        |f_{m+1}(x) + f_{m+2}(x) + \ldots + f_{n}(x)| < \varepsilon
    \end{equation*}
    whenever $n>m\geq N$ and $x \in \mathcal{I}$.
\end{theorem}
Therefore, the parameter $N$ in the uniform convergence must be valid independently of $x$, unlike the pointwise convergence. Furthermore, uniform convergence implies pointwise convergence, but not vice versa \cite{tao2016analysis}. Thus, $f_{\text{odd}}$ is neither pointwise nor uniformly convergent to $f(x)=x$ in $\mathcal{D}$. In this context, the Weierstrass M-test is a useful tool in ascertaining the uniform convergence of infinite series:
\begin{theorem}[Weierstrass M-test]
    For each $n \in \mathbb{N}$, let $f_n$ be a function defined on a set $\mathcal{I} \subseteq \mathbb{R}$, and let $M_n > 0$ be a real number satisfying     
    $$|f_n(x)| \leq M_n$$ 
    for all $x \in \mathcal{I}$. If $\sum_{n=1}^\infty M_n$ converges, then $\sum_{n=1}^\infty f_n$ converges uniformly on $\mathcal{I}$.
\end{theorem}
Applying the M-test on the series in \cref{eq:Fourier_example2} shows that $f_{\text{even}}(x)$ is uniformly convergent on $x \in \mathbb{R}$:
\begin{equation}
    \left| \dfrac{\cos[(2n-1)x]}{(2n-1)^2} \right| \leq   \dfrac{1}{(2n-1)^2}, \quad \sum_{n=1}^\infty M_n = \dfrac{\pi^2}{8},
\end{equation}
where $M_n = 1/(2n-1)^2$ was taken.

Establishing convergence criteria furnishes adequate tools for analyzing series expansions in the present unsteady aerodynamic problem. Considering the circulatory ``unit" vorticity distribution in the Wagner model \cite{epps2018vortex},
\begin{equation} \label{eq:Wagner_unity_vorticity}
    \tilde{\gamma}_b(\theta,\tau) = - R_0(\tau)   \cot{(\theta/2)}  + 2\sum_{n=1}^\infty (-1)^n R_n(\tau)\sin(n\theta),
\end{equation}
indicates a likely M-test constant in the form $M_n = R_n(\tau)$. Unfortunately, the summation $\sum_{n=1}^\infty M_n$ results in a divergent integral (via \cref{Qn_general,eq:Rn}), and no definitive statement can be made regarding the uniform convergence of \cref{eq:Wagner_unity_vorticity} through the chosen $M_n$. However, it is not necessary, once this classical formulation itself leads to the well-known discontinuity at the trailing edge. 

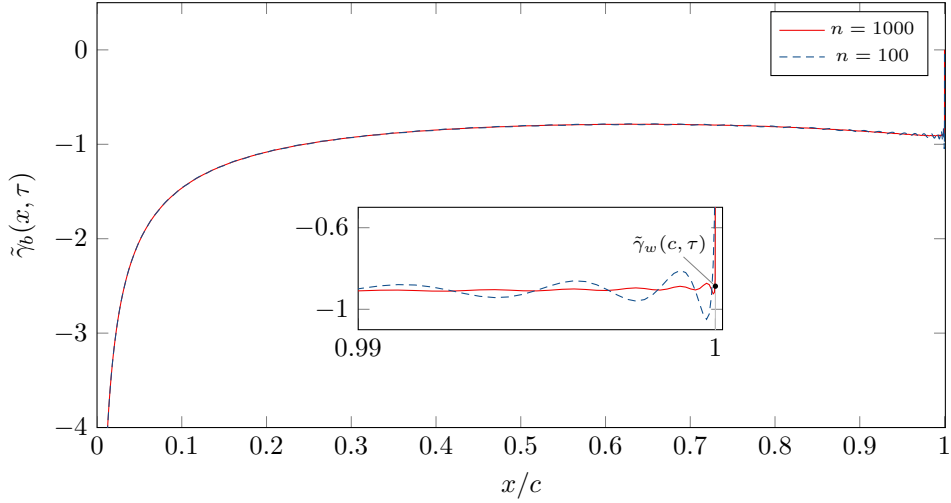
\begin{figure}[!htbp]
    \centering
    \newsavebox{\boxVorticityExample}
    \savebox{\boxVorticityExample}
    {
    \begin{tikzpicture}[>=stealth]
        \begin{axis}[
        width=0.4\linewidth,
        height=0.2\linewidth,
        xmin=0.99, xmax=1.0002,
        ymin=-1.1, ymax=-0.5,
        restrict x to domain = 0.9:1,
        xtick = {0.99,1},
        ytick = {-1,-0.6},
        legend style={font=\scriptsize},
    	]

        \draw[gray!50] (axis cs: 1,-2) -- (axis cs: 1,2);

        \addplot[Red2] table [x = x, y = g1000,col sep=comma]{table/vorticity_example.txt};

        \addplot[DodgerBlue4, densely dashed] table [x = x, y=g100,col sep=comma]{table/vorticity_example.txt};
        
        \node [circle,anchor=north,font=\footnotesize,yshift=-1.5ex] {$c+\beta\Delta x$};

        \node[circle,fill=black,minimum size=2pt,inner sep=0pt,font=\scriptsize,pin={[font=\scriptsize,pin distance=10pt,inner sep=1pt]100:$\tilde{\gamma}_w(c,\tau)$}] (c) at (axis cs: 1,-0.8867){};
    	
    	\end{axis}
    \end{tikzpicture}
    }
    \begin{tikzpicture}[>=stealth]
        \begin{axis}[
        width=0.8\linewidth,
        height=0.45\linewidth,
        xlabel={$x/c$},
        ylabel={$\Tilde{\gamma}_b(x,\tau)$},
        xmin=0, xmax=1.001,
        ymin=-4, ymax=0.5,
        restrict y to domain = -4.1:0,
        legend style={font=\scriptsize},
        clip = true,
    	]

        \addplot[Red2] table [x = x, y=g1000,col sep=comma]{table/vorticity_example.txt};
        \addlegendentry{$n=1000$}

        \addplot[DodgerBlue4,densely dashed] table [x = x, y=g100,col sep=comma]{table/vorticity_example.txt};
        \addlegendentry{$n=100$}


        \draw (axis cs: 0.485,-2.5) node{\usebox{\boxVorticityExample}};
    	
    	\end{axis}
    \end{tikzpicture}
    \caption{The circulatory ``unit'' vorticity distribution at $\tau=0.5$.}
    \label{fig:vorticity_example}
\end{figure}

As a result, several terms must be retained to visually recover vortex-sheet continuity. \cref{fig:vorticity_example} compares the circulatory ``unit'' bound-vorticity distribution computed using $n=100$ and $n=1000$ terms at time $\tau=0.5$. As shown, the continuity criterion is not mathematically satisfied, i.e., $\tilde{\gamma}_{b_{TE}}(\tau) \neq \tilde{\gamma}_{w_{TE}}(\tau)$, since the classical formulation enforces the pointwise condition $\tilde{\gamma}_b \equiv 0$ at the trailing edge. As discussed through the examples of $f_{odd}$ and $f_{even}$, such behavior suggests that the underlying series expansion is not mathematically appropriate for representing a continuous vortex-sheet distribution.

\subsubsection{Valid series expansions}
\label{sec:valid_series}

Uniformly convergent series are derived for the Wagner problem to eliminate spurious errors arising from series truncation, thereby illustrating its proper mathematical treatment. As discussed, the classical bound vorticity formulation, \cref{eq:Wagner_bound_vorticity}, does not ensure the continuity of the vortex sheet strength across the trailing edge. A possible alternative is to replace the terms $R_n(\tau)$ within a recurrence formula, similar to the procedure of \citet{sears1938systematic}. Theoretically, any coefficient $Q_n$ can be evaluated by integration by parts of \cref{Qn_general} as \cite{sears1938systematic,epps2018vortex}
\begin{equation} \label{eq:Qn_recursion}
    Q_n(k) = \dfrac{ik}{2n}\left[ Q_{n+1}(k) - Q_{n-1}(k)\right] + \dfrac{1}{n}e^{-ik},
\end{equation}
which can be replaced into \cref{eq:Rn} to provide a recurrence formula for the coefficients $R_n$:
\begin{equation} \label{eq:Rn_rec_formula}
    R_n(\tau) = \dfrac{1}{2n} \left[ R'_{n+1}(\tau) - R'_{n-1}(\tau) \right] + \dfrac{1}{n}\Psi'(\tau).
\end{equation}
\cref{eq:Rn_rec_formula} is then replaced within the classical vorticity distribution formula,
\begin{align} \label{eq:Wagner_bound_vorticity_mod0}
    \gamma_b(\theta,\tau) =& \, 2U \alpha_0 \left[ (1 - R_0(\tau) ) \dfrac{1+\cos\theta}{\sin\theta}  - \sum_{n=1}^\infty \dfrac{(-1)^n}{n} \left( R'_{n-1} - R'_{n+1} \right) \sin(n\theta) \right.  \nonumber \\
    & + \left. 2 \Psi'(\tau)\sum_{n=1}^\infty \dfrac{(-1)^n}{n} \sin(n\theta) \right],
\end{align}
where a new summation draws attention. The last summation on the right-hand side of \cref{eq:Wagner_bound_vorticity_mod0} represents the function $f(x)=-x/2$, as explored earlier in the Fourier series example given in \cref{sec:kutta_condition}, \cref{eq:Fourier_example1}:
\begin{equation} \label{eq:ex_summ_1}
    \sum_{n=1}^\infty \dfrac{(-1)^n}{n}\sin(n\theta) = -\dfrac{\theta}{2}.
\end{equation}
This series expansion is similar to $f_\text{odd}$ in \cref{eq:Fourier_example1}, which does not accurately represent $f(\theta)=\theta$ at $\theta = \pi$. The replacement of \cref{eq:ex_summ_1} into \cref{eq:Wagner_bound_vorticity_mod0} would lead to a more regular function at the trailing edge. Hence, the remaining summation in \eqref{eq:Wagner_bound_vorticity_mod0} needs particular inspection.

The proof of uniform convergence for the remaining summation is followed by taking $M_n$ as
\begin{equation} \label{eq:bound_vorticity_Mn_proof}
    \left| \frac{(-1)^n}{n} \left( R'_{n-1} - R'_{n+1} \right) \sin(n\theta) \right| \leq \frac{1}{n} \left( R'_{n-1} - R'_{n+1} \right) = M_n.
\end{equation}
Accordingly to \cref{eq:Rn,Qn_general} the series
\begin{equation} \label{eq:Mn_test}
    S = \sum_{n=1}^\infty \frac{1}{n} \left( Q_{n-1} - Q_{n+1} \right)
\end{equation}
must converge to ensure $\sum_{n=1}^\infty M_n < \infty$ in \cref{eq:bound_vorticity_Mn_proof}. 

Writing \cref{eq:Mn_test} as
\begin{equation} \label{eq:Mn_test2}
    S = Q_0 - Q_2 + \sum_{n=2}^\infty \frac{1}{n} \left( Q_{n-1} - Q_{n+1} \right)
\end{equation}
and replacing the summation
\begin{equation}
    \sum_{n=1}^\infty \dfrac{e^{-n\vartheta}}{n} = -\ln (1-e^{-\vartheta})
\end{equation}
into \cref{eq:Mn_test2} results in 
\begin{equation} \label{eq:Mn_test3}
    S = Q_0 - Q_2 - 2\int_0^\infty  e^{-ik\cosh\vartheta} \sinh\vartheta \left( e^{-\vartheta} + \ln(1 - e^{-\vartheta}) \right) \, d\vartheta,
\end{equation}
where the integral in \cref{eq:Mn_test3} is bounded, and the Weierstrass M-test is valid.

Consequently, the replacement of \cref{eq:ex_summ_1} within \cref{eq:Wagner_bound_vorticity_mod0} provides a valid series expansion for the bound vorticity in the Wagner model:
\begin{equation} \label{eq:Wagner_bound_vorticity_mod}
    \gamma_b(\theta,\tau) = 2U \alpha_0 \left[ \Phi(\tau) \cot{(\theta/2)}  + \sum_{n=1}^\infty \dfrac{(-1)^n}{n} \left( R'_{n+1} - R'_{n-1} \right) \sin(n\theta) - \Psi'(\tau)\theta \right],
\end{equation}
and the circulatory ``unit'' vorticity distribution reads
\begin{equation} \label{eq:Wagner_unity_vorticity_novel}
    \tilde{\gamma}_b(\theta,\tau) = - R_0(\tau)   \cot{(\theta/2)} + \sum_{n=1}^\infty \dfrac{(-1)^n}{n} \left( R'_{n+1} - R'_{n-1} \right) \sin(n\theta) - \Psi'(\tau)\theta.
\end{equation}
Hence, the trailing-edge bound vortex sheet strength has the form
\begin{equation}
    \gamma_b(\pi,\tau) = - 2\pi U \alpha_0 \Psi'(\tau),
\end{equation}
which ensures continuity of vorticity from \cref{eq:wake_vorticity_Wagner}.

\subsection{Discrete-vortex modeling}
\label{sec:DVM}

Careful application of numerical schemes is essential when modeling unsteady aerodynamics, particularly in discrete-vortex methods. As argued by \citet{rienstra1992note}, complications may arise when numerically evaluating series expansions such as \eqref{eq:glauert_vorticity} once the summation is necessarily finite. To illustrate that, a discrete-vortex method based on UTAT is employed to investigate the dynamics of a suddenly accelerated airfoil.

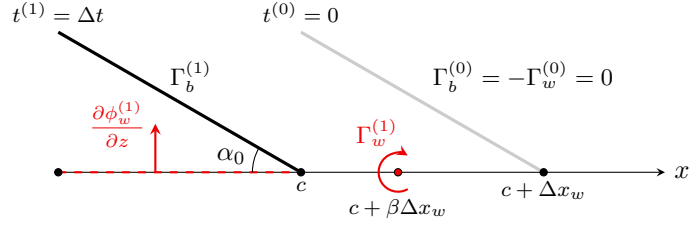
\begin{figure}[!hbtp]
    \centering
	\begin{tikzpicture}[>=stealth]
		\begin{axis}[
			axis equal,
			width=0.5\linewidth,
            height = 0.3\linewidth,
			xmin=0, xmax=1,
            ymin=0, ymax=0.5,
			axis line style={draw=none},
			ticks=none,
			clip=false,
			anchor=origin,
			]	
			
			\draw[->] (axis cs: 0,0) -- (axis cs: 1.25,0)
			 node [pos=1, anchor=west] {$x$}
			;

            \draw [black, very thick] (axis cs: 0,{0.5*tan(30)}) -- (axis cs: 1/2,0)
                node [pos=0,anchor=south,black,font=\footnotesize] {$t^{(1)}=\Delta t$}
                node [pos=0.5,anchor=south,font=\small,xshift = 1ex] {$\Gamma_b^{(1)}$}
            ;
            \draw [Red2,thick, dashed] (axis cs: 0,0) -- (axis cs: 1/2,0);

            \draw [Red2,thick,->] (axis cs: 0.2,0) -- (axis cs: 0.2,0.1)
            node [pos=1,anchor= east,font=\scriptsize] {$\dfrac{\partial \phi_w^{(1)}}{\partial z}$}
            ;

            \draw [black!20, very thick] (axis cs: 0.5,{0.5*tan(30)}) -- (axis cs: 1,0)
            node [pos=0,anchor=south,black,font=\footnotesize] {$t^{(0)}=0$}
            node [pos=0.5,anchor=south west,font=\small,black] {$\Gamma_b^{(0)} = - \Gamma_w^{(0)} = 0$}
            ;

            \draw[] (axis cs: 1/2,0) + (180:10) arc (180:150:10)
                node [pos=0.65,anchor=east,black] {$\alpha_0$}
            ;

            \draw [fill=black] (axis cs: 0,0) circle [radius=0.0075];
            \draw [fill=black] (axis cs: 1/2,0) circle [radius=0.0075]
                node [anchor=north,font=\footnotesize] {$c$}
            ;
            \draw [fill=black] (axis cs: 1,0) circle [radius=0.0075]
                node [anchor=north,font=\footnotesize] {$c+\Delta x_w$}
            ;

            \draw[Red2,thick,->] (axis cs: 0.7,0) + (-60:4) arc (-60:-300:4)
                node [pos=1,anchor=south east,font=\small] {$\Gamma_w^{(1)}$}
            ;

            \draw [fill=Red2] (axis cs: 0.7,0) circle [radius=0.0075]
                node [anchor=north,font=\footnotesize,yshift=-1.5ex] {$c+\beta\Delta x_w$}
            ;

		\end{axis}
	\end{tikzpicture}
	\caption{Discrete-vortex method employed to analyze a suddenly accelerated airfoil.}
	\label{fig:DVM_model}
\end{figure}

\cref{fig:DVM_model} depicts a flat plate at $t=0$ and $t=\Delta t$, where $\Delta t$ is the time step. The discrete-vortex method in this case is built upon the same framework as thin airfoil theory: considering small-disturbance approximations, where the wake vorticity is only convected with the freestream velocity, i.e, $\Delta x_w = U \Delta t$, chordwise. Thus, this vortex sheet element is replaced by a point vortex located at $x=c + \beta \Delta x_w$, where $\beta \in (0,1)$ denotes a positioning parameter within the continuous sheet element $\Delta x_w$. \citet{katz2001low} argued that a typical value for $\beta$ would be in the range $0.2 - 0.3$, although $\beta = 0.5$ is taken for other authors. In any case, the chosen value should recover the analytical results as $\Delta t \to 0$.

The following calculations describe the analysis of the initial shed vortex, with subsequent vortices analyzed accordingly. As shown in \cref{fig:DVM_model}, the first wake vortex is placed at $x_w = c+\beta U \Delta t$, or in its dimensionless form, $x_w/c =   1 + \beta \Delta \tau /2 $, where $\Delta \tau = (2U/c) \Delta t$. Making use of transforms \eqref{eq:Glauert_transform} and
\begin{equation} \label{eq:wake_variable_transform}
    x_w = \dfrac{c}{2}(1 + \cosh\vartheta), \quad \vartheta \in [0,\infty),
\end{equation}
the induced velocity at the airfoil surface has the form
\begin{equation} \label{eq:DVM_induced_velocity}
    \dfrac{\partial \phi_w^{(1)}}{\partial z} = \dfrac{\Gamma_w^{(1)}}{\pi c(\cos\theta + \cosh\vartheta_1)},
\end{equation}
where $\cosh\vartheta_m = 1 + (\beta + m - 1) \Delta \tau$ refers to the location of the $m$-th vortex, and the superscript indicates the time step, i.e., $\{ f\}^{(j)} \equiv f(j\Delta \tau)$ for any arbitrary function $f$ and index $j=0,1,2,\ldots$ The integral \eqref{eq:DVM_integral_final} is invoked and the Fourier coefficients read
\begin{align}
    A_0^{(1)} &=   \alpha_0 + \dfrac{1}{\sinh\vartheta_1}\left(\dfrac{\Gamma_w^{(1)}}{\pi c U}\right), \label{eq:A0_DVM0} \\
    A_n^{(1)} &= - 2(-1)^n\dfrac{e^{-n\vartheta_1}}{\sinh\vartheta_1}\left(\dfrac{\Gamma_w^{(1)}}{\pi c U}\right),  \quad n=1,2,3,\ldots \label{eq:An_DVM0}
\end{align}

From Kelvin's circulation theorem \eqref{eq:Kelvin_theorem}, $\Gamma_b^{(1)} = - \Gamma_w^{(1)}$, and the coefficients needed to compute the lift coefficient will have the form
\begin{align}
    A_0^{(1)} &=   \alpha_0\dfrac{\cosh\vartheta_1}{1+\cosh\vartheta_1}, \label{eq:A0_DVM} \\
    A_n^{(1)} &= 2(-1)^n\alpha_0 \dfrac{e^{-n\vartheta_1}}{1+\cosh\vartheta_1}, \label{eq:An_DVM}
\end{align}
where the bound circulation reads
\begin{equation} \label{eq:DVM_Wagner_Gamma}
    \Gamma_b^{(1)} = \pi c U \alpha_0 \tanh(\vartheta_1/2).
\end{equation}

In the limit $\Delta \tau \to 0$, one obtains $\vartheta_1 \to 0$, ensuring zero bound circulation, as expected. Furthermore, the wake vorticity computed at the trailing edge can be approximated as
\begin{equation}
    \left.\gamma^{(1)}_w\right|_{\vartheta_1=0} \approx \dfrac{\Gamma_w^{(1)}}{\Delta x_w} = -\dfrac{2\Gamma_b^{(1)}}{c\Delta \tau}.
\end{equation}

The derivatives $A'_n$ are computed using a backward finite differences method,
\begin{equation}
    A_n^{'(j)} \approx \dfrac{A_n^{(j)} - A_n^{(j-1)}}{\Delta \tau}, \quad j=1,2,3,\ldots
\end{equation}
by recalling $A_n^{(0)} \equiv 0$, i.e., assuming fluid initially at rest. Therefore, in the limit $\Delta \tau \to 0$:
\begin{equation} \label{eq:gamma_wake_DVM}
    \left.\gamma^{(1)}_w\right|_{\vartheta_1=0} \approx -\dfrac{{\Gamma_b'}^{(1)}}{c/2},
\end{equation}
which recovers \cref{eq:rate_vorticity2}. Thus, at the first time step, the bound circulation and trailing-edge wake vorticity obtained from the discrete-vortex method are consistent with Wagner’s results, \cref{eq:initial_value_bound_circ_wake_vorticity}.

However, the trailing-edge bound vorticity computed through the classical \cref{eq:glauert_vorticity} is zero, implying that the vorticity field is not mathematically continuous across the trailing edge. In addition, from \cref{eq:pressure_katz2}, the trailing-edge pressure difference is given by
\begin{equation}
    \left.\Delta C_p^{(1)}\right|_{\theta=\pi} = 4\pi \left( {A'_0}^{(1)} + \dfrac{{A'_1}^{(1)}}{2} \right),
\end{equation}
which is nonzero in unsteady motions, contradicting the condition of zero loading at the trailing edge. Actually, not only is it nonzero, but it also diverges as $\Delta \tau \to 0$.

To obtain numerically smooth solutions, selected time derivatives may be replaced by equivalent expressions (e.g., through recurrence formulae or consistent governing equations) in order to improve regularity. When the unsteady Kutta condition is correctly enforced, the resulting formulation yields numerically stable solutions.

\subsubsection{Valid series expansions}
\label{sec:valid_series_DVM}

Consistent with the previously developed analytical framework, a recurrence relation for the coefficients $A_n$ in the Wagner model is derived. Firstly, the function $\Psi'(\tau)$ is written in terms of $A_0'$ and $A_1'$ from \cref{eq:wagner_gamma,eq:airfoil_circulation}:
\begin{equation} \label{eq:rel_Psi_An}
    \Psi'(\tau) = \dfrac{1}{\alpha_0} \left( A'_0 + \dfrac{A'_1}{2} \right).
\end{equation}

Therefore, the recurrence formula for $A_n$ is obtained by isolating the coefficients in \cref{eq:An_Wagner},
\begin{equation} \label{eq:An_Wagner_Rn}
    R_0(\tau) = 1 - \dfrac{A_0}{\alpha_0}; \qquad R_n(\tau) = \dfrac{(-1)^n}{2\alpha_0}  A_n, \quad n\geq 1,
\end{equation}
and replacing them within the recurrence formula for $R_n$ in \eqref{eq:Rn_rec_formula}:
\begin{equation}
    A_n(\tau) = \dfrac{1}{2n} \left( A'_{n-1} - A'_{n+1} \right) + 2\dfrac{(-1)^n}{n}\left( A'_0 + \dfrac{A'_1}{2} \right), \quad n \geq 2.
\end{equation}
Computing for $A_1$ requires careful analysis. Taking $n=1$ in \cref{eq:Rn_rec_formula}:
\begin{equation}
    R_1 = \dfrac{1}{2}(R'_2 - R'_0) + \Psi',
\end{equation}
which leads to the recurrence formula for $A_1$ through \cref{eq:rel_Psi_An,eq:An_Wagner_Rn}:
\begin{equation} \label{eq:An_recurrence_n1}
    A_1 = -3A'_0 - A'_1 - \dfrac{A'_2}{2}.
\end{equation}

The bound vorticity formulation in the DVM framework is given by
\begin{equation} \label{eq:bound_vorticity_DVM1}
    \dfrac{\gamma_b(\theta,\tau)}{2U} = A_0 \cot{(\theta/2)} + A_1\sin\theta + \sum_{n=2}^\infty \left[ \dfrac{1}{2n} \left( A'_{n-1} - A'_{n+1} \right) + 2\dfrac{(-1)^n}{n}\left( A'_0 + \dfrac{A'_1}{2} \right) \right] \sin(n\theta),
\end{equation}
which can be rewritten as
\begin{align} \label{eq:bound_vorticity_DVM2}
    \dfrac{\gamma_b(\theta,\tau)}{2U} =& \, A_0 \cot{(\theta/2)} + \left(A_1 + 2A'_0 + A'_1 +\dfrac{A'_2}{2} \right)\sin\theta +\dfrac{A'_1}{4}\sin(2\theta) + 2 \left( A'_0 + \dfrac{A'_1}{2} \right) \sum_{n=1}^\infty \dfrac{(-1)^n}{n}\sin(n\theta)  \nonumber \\
    &- \sum_{n=2}^\infty\dfrac{A'_n}{2} \left( \dfrac{\sin[(n-1)\theta]}{n-1} - \dfrac{\sin[(n+1)\theta]}{n+1} \right).
\end{align}
From the series expansion \eqref{eq:ex_summ_1} and \cref{eq:An_recurrence_n1}:
\begin{align} \label{eq:bound_vorticity_DVM}
    \dfrac{\gamma_b(\theta,\tau)}{2U} =& \, A_0 \cot{(\theta/2)} - A'_0 \sin\theta +\dfrac{A'_1}{4}\sin(2\theta) - \left( A'_0 + \dfrac{A'_1}{2} \right) \theta  \nonumber \\
    &- \sum_{n=2}^\infty\dfrac{A'_n}{2} \left( \dfrac{\sin[(n-1)\theta]}{n-1} - \dfrac{\sin[(n+1)\theta]}{n+1} \right).
\end{align}
Hence, at the trailing edge:
\begin{equation}
    \gamma_b(\pi,\tau) = - 2\pi U \left( A'_0 + \dfrac{A'_1}{2} \right) = -\dfrac{2}{c} \Gamma'_b(\tau),
\end{equation}
which along with \cref{eq:rate_vorticity2} ensures vorticity continuity.

The pressure difference at the trailing edge is regularized accordingly. From \cref{eq:pressure_katz2}:
\begin{align} \label{eq:pressure_katz_DVM}
	\dfrac{\Delta C_p(\theta,\tau)}{2/U}  &=    \gamma_b(\theta,\tau) +  2U \left[  \left( A'_0 + \dfrac{A'_1}{2} \right) \theta + A'_0 \sin\theta - \dfrac{A'_1}{4} \sin(2\theta) \right. \nonumber \\
    &+ \left. \sum_{n=2}^\infty \dfrac{A'_n}{2} \left( \dfrac{\sin{\left[(n-1)\theta \right]}}{n-1}  - \dfrac{\sin{\left[(n+1)\theta \right]}}{n+1}\right) \right],
\end{align}
which is replaced within the uniformly valid formulation \eqref{eq:bound_vorticity_DVM}:
\begin{align} \label{eq:pressure_katz_DVM2}
	\Delta C_p(\theta,\tau) = 4A_0(\tau) \cot(\theta/2).
\end{align}
The new distribution equation ensures zero loading at the trailing edge, therefore satisfying the unsteady Kutta condition. 

Finally, the lift coefficient is obtained by direct integration of \eqref{eq:pressure_katz_DVM2} over the chord. However, to demonstrate the internal consistency of the proposed framework, algebraic manipulations based on the recurrence relation are performed, leading to a valid representation. Rearranging the lift coefficient \eqref{eq:normal_force}:
\begin{equation} \label{eq:normal_force_DVM1}
    C_l(\tau) = 2 \pi A_0 + \pi \left( A_1 + 3A'_0 + A'_1 + \dfrac{A'_2}{2} \right),
\end{equation}
which from relation \eqref{eq:An_recurrence_n1}:
\begin{equation} \label{eq:normal_force_DVM2}
    C_l(\tau) = 2 \pi A_0.
\end{equation}
Consistently, the replacement of $A_0(\tau)=\alpha_0 \Phi(\tau)$ from the step-input model into pressure distribution \eqref{eq:pressure_katz_DVM2} and lift coefficient \eqref{eq:normal_force_DVM2} recovers the classical results of Wagner as presented in \cref{sec:Wagner}.

\section{Results and Validation}
\label{sec:results}

This section evaluates the proposed bound-vorticity formulation for the Wagner problem through direct comparisons with the classical series representation. The analysis focuses on two central aspects: (i) the unsteady Kutta condition at the trailing edge and (ii) the implications of the modified formulation for discrete-vortex modeling and aerodynamic load prediction. Particular attention is given to the continuity of the vortex-sheet strength and the regularization of the pressure distribution near the trailing edge. The proposed analytical framework is validated against the exact Wagner solution and compared with the classical unsteady thin-airfoil formulation.

\subsection{Continuous vortex-sheet formulation}

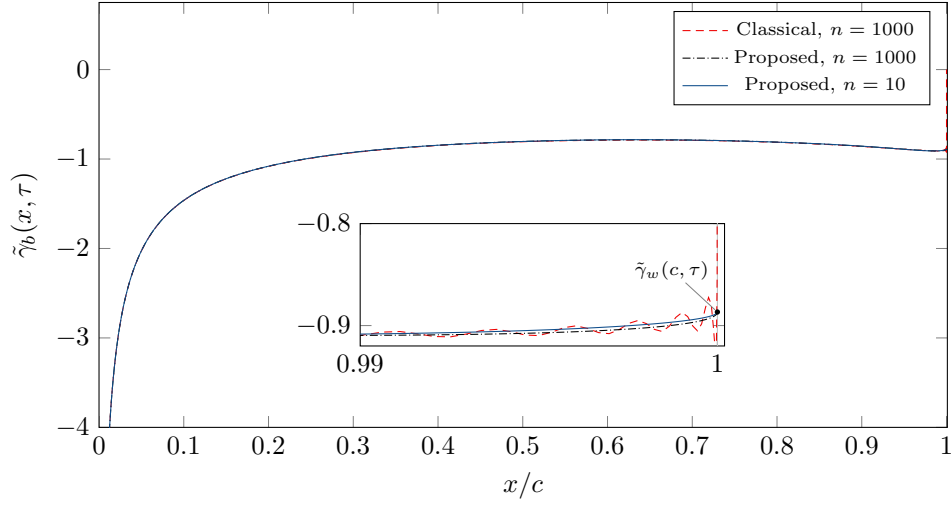
\begin{figure}[!htbp]
    \centering
    \newsavebox{\boxVorticityMod}
    \savebox{\boxVorticityMod}
    {
    \begin{tikzpicture}[>=stealth]
    	\begin{axis}[
    	width=0.4\linewidth,
    	height=0.2\linewidth,
    	xmin=0.99, xmax=1.0002,
        ymin=-0.92, ymax=-0.8,
        restrict x to domain = 0.9:1,
        xtick = {0.99,1},
        ytick = {-0.9,-0.8},
        legend style={font=\scriptsize},
    	]

        \draw[gray!50] (axis cs: 1,-2) -- (axis cs: 1,2);

        \addplot[densely dashed, Red2] table [x = x, y = g1000,col sep=comma]{table/vorticity_example.txt};

        \addplot[densely dashdotted] table [x = x, y = g1000m,col sep=comma]{table/vorticity_example.txt};

        \addplot[DodgerBlue4] table [x = x, y=g10m,col sep=comma]{table/vorticity_example.txt};
        
        \node [circle,anchor=north,font=\footnotesize,yshift=-1.5ex] {$c+\beta\Delta x$};

        \node[circle,fill=black,minimum size=2pt,inner sep=0pt,font=\scriptsize,pin={[font=\scriptsize,pin distance=10pt,inner sep=1pt]100:$\tilde{\gamma}_w(c,\tau)$}] (c) at (axis cs: 1,-0.8867){}; 
    	
    	\end{axis}
    \end{tikzpicture}
    }
    \begin{tikzpicture}[>=stealth]
    	\begin{axis}[
    	width=0.8\linewidth,
    	height=0.45\linewidth,
        xlabel={$x/c$},
    	ylabel={$\Tilde{\gamma}_b(x,\tau)$},
    	xmin=0, xmax=1.001,
        ymin=-4, ymax=0.75,
        restrict y to domain = -8:8,
        legend style={font=\scriptsize},
        clip = true,
    	]

        \addplot[densely dashed, Red2] table [x = x, y = g1000,col sep=comma]{table/vorticity_example.txt};
        \addlegendentry{Classical, $n=1000$}

        \addplot[densely dashdotted] table [x = x, y=g1000m,col sep=comma]{table/vorticity_example.txt};
        \addlegendentry{Proposed, $n=1000$}

        \addplot[DodgerBlue4] table [x = x, y=g10m,col sep=comma]{table/vorticity_example.txt};
        \addlegendentry{Proposed, $n=10$}


        \draw (axis cs: 0.485,-2.5) node{\usebox{\boxVorticityMod}};
    	
    	\end{axis}
    \end{tikzpicture}
    \caption{Comparison of the circulatory ``unit" vorticity distribution at $\tau=0.5$ computed from the classical \eqref{eq:Wagner_unity_vorticity} and proposed \eqref{eq:Wagner_unity_vorticity_novel} formulations.}
    \label{fig:vorticity_mod}
\end{figure}

\cref{fig:vorticity_mod} compares the circulatory ``unit'' bound-vorticity distribution computed from the classical formulation, \cref{eq:Wagner_unity_vorticity}, and from the proposed uniformly convergent expansion, \cref{eq:Wagner_unity_vorticity_novel}, at $\tau=0.5$. The classical expansion requires a large number of terms to visually recover continuity near the trailing edge, yet the vorticity remains mathematically discontinuous since $\tilde{\gamma}_{b_{TE}}(\tau)\neq\tilde{\gamma}_{w_{TE}}(\tau)$. In contrast, the proposed formulation converges rapidly and recovers the correct trailing-edge limit with only a few summation terms. The oscillatory behavior associated with the classical expansion is eliminated, leading to a smooth vortex-sheet distribution over the entire chord.

The differences become more evident near the trailing edge, where the classical representation imposes the pointwise condition $\tilde{\gamma}_b(\pi,\tau)\equiv0$. Although this condition is mathematically acceptable in steady flows, it becomes inconsistent in unsteady motions because the shed vorticity at the trailing edge is generally nonzero. Consequently, the classical formulation introduces an artificial discontinuity in the vortex-sheet strength. The proposed series expansion resolves this inconsistency by enforcing
\begin{equation}
    \tilde{\gamma}_{b_{TE}}(\tau) = \tilde{\gamma}_{w_{TE}}(\tau)
\end{equation}
thereby satisfying the continuity requirement associated with the unsteady Kutta condition.

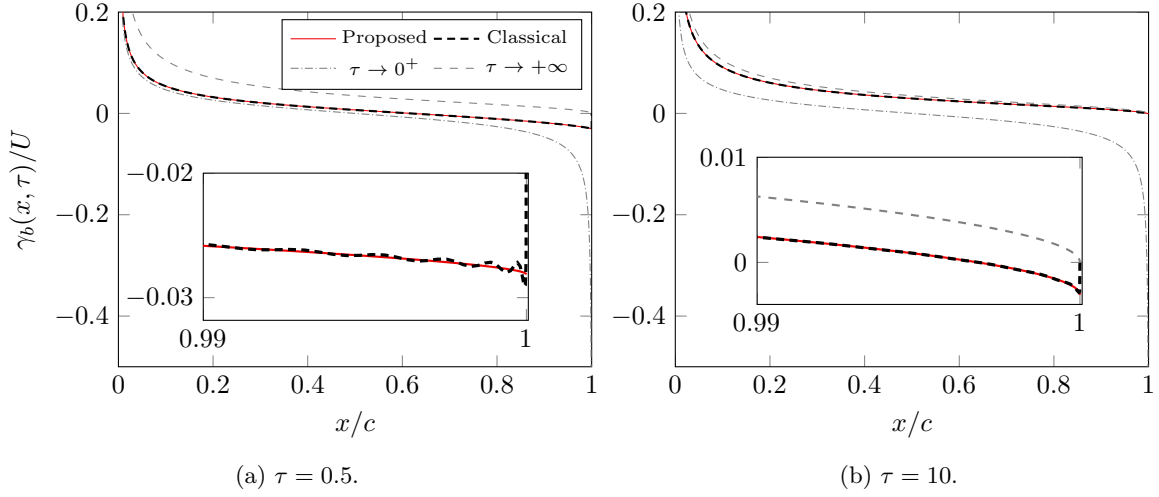
\begin{figure}[!htbp]
\centering
 \begin{subfigure}[b]{.49\textwidth}
    \newsavebox{\boxWagnerOne}
    \savebox{\boxWagnerOne}
    {\begin{tikzpicture}[>=stealth]
    	\begin{axis}[
    	width=0.75\linewidth,
    	height=0.45\linewidth,
    	xmin=0.991, xmax=1.000055,
        ymin=-0.035, ymax=-0.022,
        xtick = {0.991, 1},
        ytick = {-0.033,-0.022},
        xticklabels={$0.99$,$1$},
        yticklabels={$-0.03$,$-0.02$},
        legend style={font=\scriptsize},
        clip = true,
        restrict y to domain = -1:1,
    	]

        \addplot[Red2, thick] table [x = x, y = gamma_mod_t1, col sep=comma] {table/bound_vorticity_Wagner.txt};

        \addplot[densely dashed, very thick] table [x = x, y = gamma_t1, col sep=comma] {table/bound_vorticity_Wagner.txt};

        \addplot[gray, densely dashdotted] table [x = x, y = lim_zero, col sep=comma] {table/bound_vorticity_Wagner.txt};

        \addplot[gray, thick, dashed] table [x = x, y = lim_inf, col sep=comma] {table/bound_vorticity_Wagner.txt};

    	\end{axis}
    \end{tikzpicture}}
    \begin{tikzpicture}[>=stealth]
    	\begin{axis}[
    	width=\linewidth,
    	height=0.8\linewidth,
        xlabel={$x/c$},
    	ylabel={$\gamma_b(x,\tau)/U$},
    	xmin=0, xmax=1,
        ymin=-0.5, ymax=0.2,
        legend columns = 2,
        legend style={font=\scriptsize},
        clip = true,
        restrict y to domain = -1:1,
    	]

        \addplot[Red2] table [x = x, y = gamma_mod_t1, col sep=comma] {table/bound_vorticity_Wagner.txt};
        \addlegendentry{Proposed}

        \addplot[densely dashed, thick] table [x = x, y = gamma_t1, col sep=comma] {table/bound_vorticity_Wagner.txt};
        \addlegendentry{Classical}

        \addplot[gray, densely dashdotted] table [x = x, y = lim_zero, col sep=comma] {table/bound_vorticity_Wagner.txt};
        \addlegendentry{$\tau \rightarrow 0^+$}

        \addplot[gray, dashed] table [x = x, y = lim_inf, col sep=comma] {table/bound_vorticity_Wagner.txt};
        \addlegendentry{$\tau \rightarrow +\infty$}

        \draw (axis cs: 0.445,-0.28)
        node{\usebox{\boxWagnerOne}};
    	
    	\end{axis}
    \end{tikzpicture}
 \caption{$\tau = 0.5$.}
 \end{subfigure}
\begin{subfigure}[b]{.49\textwidth}
    \newsavebox{\boxWagnerTwo}
    \savebox{\boxWagnerTwo}
    {\begin{tikzpicture}[>=stealth]
    	\begin{axis}[
    	width=0.75\linewidth,
    	height=0.45\linewidth,
    	xmin=0.992, xmax=1.000055,
        ymin=-0.002, ymax=0.005,
        xtick = {0.992, 1},
        ytick = {0,0.005},
        yticklabels={$0$,$0.01$},
        xticklabels={$0.99$,$1$},
        legend style={font=\scriptsize},
        clip = true,
        restrict y to domain = -1:1,
    	]

        \addplot[Red2, thick] table [x = x, y = gamma_mod_t2, col sep=comma] {table/bound_vorticity_Wagner.txt};

        \addplot[densely dashed, very thick] table [x = x, y = gamma_t2, col sep=comma] {table/bound_vorticity_Wagner.txt};

        \addplot[gray, densely dashdotted] table [x = x, y = lim_zero, col sep=comma] {table/bound_vorticity_Wagner.txt};

        \addplot[gray, thick, dashed] table [x = x, y = lim_inf, col sep=comma] {table/bound_vorticity_Wagner.txt};

    	\end{axis}
    \end{tikzpicture}}
    \begin{tikzpicture}[>=stealth]
    	\begin{axis}[
    	width=\linewidth,
    	height=0.8\linewidth,
        xlabel={$x/c$},
    	xmin=0, xmax=1,
        ymin=-0.5, ymax=0.2,
        legend style={font=\scriptsize},
        clip = true,
        restrict y to domain = -1:1,
    	]

        \addplot[Red2] table [x = x, y = gamma_mod_t2, col sep=comma] {table/bound_vorticity_Wagner.txt};

        \addplot[densely dashed, thick] table [x = x, y = gamma_t2, col sep=comma] {table/bound_vorticity_Wagner.txt};

        \addplot[gray, densely dashdotted] table [x = x, y = lim_zero, col sep=comma] {table/bound_vorticity_Wagner.txt} ;

        \addplot[gray, dashed] table [x = x, y = lim_inf, col sep=comma] {table/bound_vorticity_Wagner.txt};

        \draw (axis cs: 0.46,-0.25) node{\usebox{\boxWagnerTwo}};
    	
    	\end{axis}
    \end{tikzpicture}
 \caption{$\tau = 10$.}
 \end{subfigure}
 \caption{Comparison of the bound vorticity distribution computed from the classical \eqref{eq:Wagner_bound_vorticity} and proposed \eqref{eq:Wagner_bound_vorticity_mod} formulations.}
 \label{fig:bound_vorticity_Wagner}
\end{figure}

\cref{fig:bound_vorticity_Wagner} compares the bound-vorticity distributions at $\tau=0.5$ and $\tau=10$. At small times, the classical expansion exhibits pronounced oscillations near the trailing edge due to the contribution of the bound-circulation rate. As $\tau\rightarrow\infty$, both formulations converge toward the steady-state solution, and the discrepancies become negligible. This behavior indicates that the mathematical inconsistency of the classical formulation is primarily associated with transient unsteady effects.

\subsection{Discrete-vortex modeling}

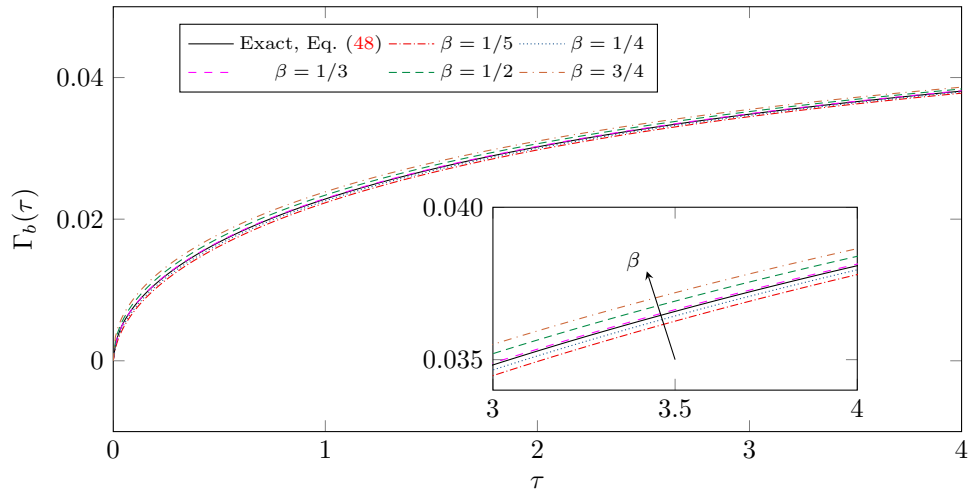
\begin{figure}[!htbp]
\centering
    \newsavebox{\boxCirculationUTAT}
    \savebox{\boxCirculationUTAT}
    {\begin{tikzpicture}[>=stealth]
    	\begin{axis}[
    	width=0.4\linewidth,
    	height=0.25\linewidth,
    	xmin=3, xmax=4,
        ymin=0.034, ymax=0.04,
        xtick = {3,3.5,4},
        ytick = {0.035,0.04},
        yticklabels={$0.035$,$0.040$},
        legend style={font=\scriptsize},
        clip = true,
        restrict y to domain = -1:1,
    	]

        \addplot[] table [x = x, y = Gamma_Wagner, col sep=comma] {table/Gamma_UTAT_beta.txt};

        \addplot[Red2, densely dashdotted] table [x = x, y = one5, col sep=comma] {table/Gamma_UTAT_beta.txt};

        \addplot[DodgerBlue4,densely dotted] table [x = x, y = one4, col sep=comma] {table/Gamma_UTAT_beta.txt};

        \addplot[Magenta1, dashed] table [x = x, y = one3, col sep=comma] {table/Gamma_UTAT_beta.txt};

        \addplot[SpringGreen4, densely dashed] table [x = x, y = one2, col sep=comma] {table/Gamma_UTAT_beta.txt};

        \addplot[Sienna3, dashdotted] table [x = x, y = one1, col sep=comma] {table/Gamma_UTAT_beta.txt};

        \draw[->] (axis cs: 3.5,0.035) --+ (105:30)
        node [pos=0.9,anchor=south east,font=\footnotesize] {$\beta$}
        ;

    	\end{axis}
    \end{tikzpicture}}
    \begin{tikzpicture}[>=stealth]
    	\begin{axis}[
    	width=0.8\linewidth,
    	height=0.45\linewidth,
        xlabel={$\tau$},
    	ylabel={$\Gamma_b(\tau)$},
    	xmin=0, xmax=4,
        ymin=-0.01, ymax=0.05,
        xtick = {0,1,...,4},
        ytick = {0,0.02,0.04},
        yticklabels={$0$,$0.02$,$0.04$},
        legend columns = 3,
        legend style={font=\scriptsize, at={(0.36,0.96)}, anchor=north},
        clip = true,
        restrict x to domain = 0:4,
    	]

        \addplot[] table [x = x, y = Gamma_Wagner, col sep=comma] {table/Gamma_UTAT_beta.txt};
        \addlegendentry{Exact, \cref{eq:wagner_gamma}}

        \addplot[Red2, densely dashdotted] table [x = x, y = one5, col sep=comma] {table/Gamma_UTAT_beta.txt};
        \addlegendentry{$\beta = 1/5$}

        \addplot[DodgerBlue4,densely dotted] table [x = x, y = one4, col sep=comma] {table/Gamma_UTAT_beta.txt};
        \addlegendentry{$\beta = 1/4$}

        \addplot[Magenta1, dashed] table [x = x, y = one3, col sep=comma] {table/Gamma_UTAT_beta.txt};
        \addlegendentry{$\beta = 1/3$}

        \addplot[SpringGreen4, densely dashed] table [x = x, y = one2, col sep=comma] {table/Gamma_UTAT_beta.txt};
        \addlegendentry{$\beta = 1/2$}

        \addplot[Sienna3, dashdotted] table [x = x, y = one1, col sep=comma] {table/Gamma_UTAT_beta.txt};
        \addlegendentry{$\beta = 3/4$}

        \draw (axis cs: 2.5,0.0075)         node{\usebox{\boxCirculationUTAT}};        
    	
    	\end{axis}
    \end{tikzpicture}
 \caption{Effect of the parameter $\beta$ on the computed bound circulation in discrete-vortex modeling. The arrow points in the direction of increasing $\beta$ values.}
 \label{fig:Gamma_UTAT_beta}
\end{figure}

The discrete-vortex method developed in \cref{sec:DVM} is now employed to investigate the implications of the proposed formulation in numerical simulations. \cref{fig:Gamma_UTAT_beta} presents the effect of the parameter $\beta$ on the computed bound circulation. As expected, the solution converges toward Wagner's analytical result as $\Delta\tau\rightarrow0$. Among the tested values, $\beta=1/3$ provides the best overall agreement with the exact solution for a given time step, suggesting that this value more accurately represents the equivalent position of a continuous vortex-sheet element through a single discrete vortex.

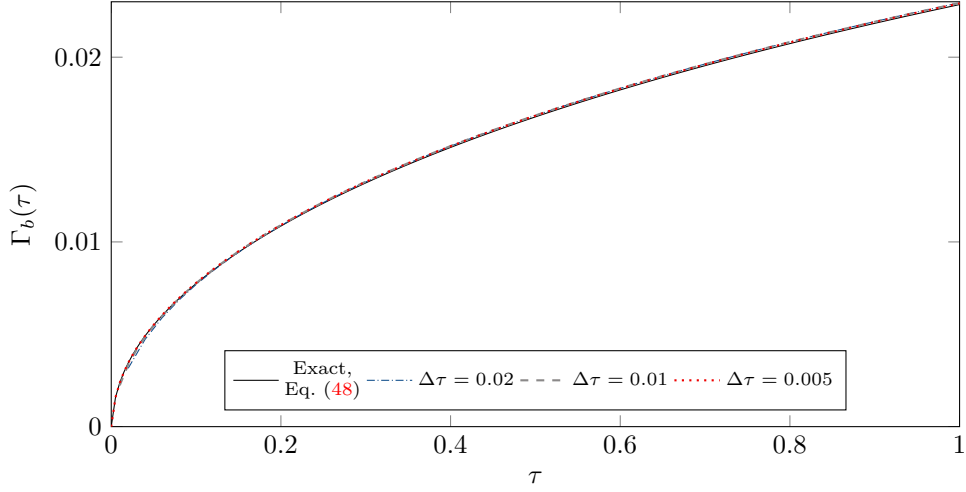
\begin{figure}[!htbp]
\centering
    \begin{tikzpicture}[>=stealth]
    	\begin{axis}[
    	width=0.8\linewidth,
    	height=0.45\linewidth,
        xlabel={$\tau$},
    	ylabel={$\Gamma_b(\tau)$},
    	xmin=0, xmax=1,
        ymin=0, ymax=0.023,
        xtick = {0,0.2,...,1},
        ytick = {0,0.01,0.02},
        yticklabels={$0$,$0.01$,$0.02$,$0.03$},
        legend columns = 4,
        legend style={font=\scriptsize, at={(0.5,0.18)}, anchor=north, cells={align=center}},
        clip = true,
        restrict x to domain = 0:4,
    	]

        \addplot[] table [x = x, y = Gamma_Wagner, col sep=comma] {table/Gamma_UTAT_dt3.txt};
        \addlegendentry{Exact, \\ \cref{eq:wagner_gamma}}

        \addplot[DodgerBlue4, densely dashdotted] table [x = x, y = Gamma, col sep=comma] {table/Gamma_UTAT_dt1.txt};
        \addlegendentry{$\Delta\tau = 0.02$}

        \addplot[Snow4, dashed, thick] table [x = x, y = Gamma, col sep=comma] {table/Gamma_UTAT_dt2.txt};
        \addlegendentry{$\Delta\tau = 0.01$}

        \addplot[Red2, dotted, thick] table [x = x, y = Gamma, col sep=comma] {table/Gamma_UTAT_dt3.txt};
        \addlegendentry{$\Delta\tau = 0.005$}
    	
    	\end{axis}
    \end{tikzpicture}
 \caption{Comparison of the bound circulation computed for different time steps through the discrete-vortex model.}
 \label{fig:Gamma_UTAT_dt}
\end{figure}

The convergence behavior with respect to the time step is shown in \cref{fig:Gamma_UTAT_dt}. Reducing $\Delta\tau$ improves the agreement between the discrete-vortex model and the analytical solution, confirming the consistency of the numerical implementation. Even though both the classical and proposed formulations recover the same bound-circulation history, substantial differences arise when evaluating local aerodynamic quantities such as the bound-vorticity and pressure distributions.

\begin{figure}[!htbp]
\centering
    \newsavebox{\boxBoundVorticityUTAT}
    \savebox{\boxBoundVorticityUTAT}
    {\begin{tikzpicture}[>=stealth]
    	\begin{axis}[
    	width=0.45\linewidth,
    	height=0.25\linewidth,
    	xmin=0.95, xmax=1,
        ymin=-0.032, ymax=-0.02,
        xtick = {0.95, 1},
        ytick = {-0.03,-0.02},
        yticklabels={$-0.03$,$-0.02$},
        legend style={font=\scriptsize},
        clip = true,
        restrict y to domain = -1:1,
    	]

        \addplot[] table [x = x, y = gamma_mod_t1, col sep=comma] {table/bound_vorticity_Wagner.txt};

        \addplot[DodgerBlue4,densely dashdotted, thick] table [x = x, y = gamma, col sep=comma] {table/bound_vorticity_UTAT.txt};

        \addplot[Red2,densely dashed, thick] table [x = x, y = gamma_mod, col sep=comma] {table/bound_vorticity_UTAT.txt};

        \node[circle,fill=black,minimum size=2pt,inner sep=0pt,font=\scriptsize,pin={[font=\scriptsize,pin distance=10pt,inner sep=1pt]180:$\gamma_w(c,\tau)$}] (c) at (axis cs: 1,-0.030952){};

    	\end{axis}
    \end{tikzpicture}}
    \begin{tikzpicture}[>=stealth]
    	\begin{axis}[
    	width=0.8\linewidth,
    	height=0.45\linewidth,
        xlabel={$x/c$},
    	ylabel={$\gamma_b(x,\tau)/U$},
    	xmin=0, xmax=1,
        ymin=-0.22, ymax=0.12,
        xtick = {0,0.2,...,1},
        ytick = {-0.2,-0.1,...,0.1},
        legend columns = 5,
        legend style={font=\scriptsize, at={(0.5,0.98)}, anchor=north, cells={align=center}},
        clip = true,
        restrict y to domain = -1:1,
    	]

        \addplot[] table [x = x, y = gamma_mod_t1, col sep=comma] {table/bound_vorticity_Wagner.txt};
        \addlegendentry{Exact, \\ \cref{eq:Wagner_bound_vorticity_mod}}

        \addplot[DodgerBlue4,densely dashdotted, thick] table [x = x, y = gamma, col sep=comma] {table/bound_vorticity_UTAT.txt};
        \addlegendentry{Classical, \\ \cref{eq:glauert_vorticity}}

        \addplot[Red2,densely dashed, thick] table [x = x, y = gamma_mod, col sep=comma] {table/bound_vorticity_UTAT.txt};
        \addlegendentry{Proposed, \\ \cref{eq:bound_vorticity_DVM}}
        
        \draw (axis cs: 0.445,-0.115)         node{\usebox{\boxBoundVorticityUTAT}};
    	
    	\end{axis}
    \end{tikzpicture}
 \caption{Comparison of the bound vorticity distribution at $\tau=0.5$ computed through the classical and proposed formulations in the discrete-vortex model.}
 \label{fig:bound_vorticity_UTAT}
\end{figure}
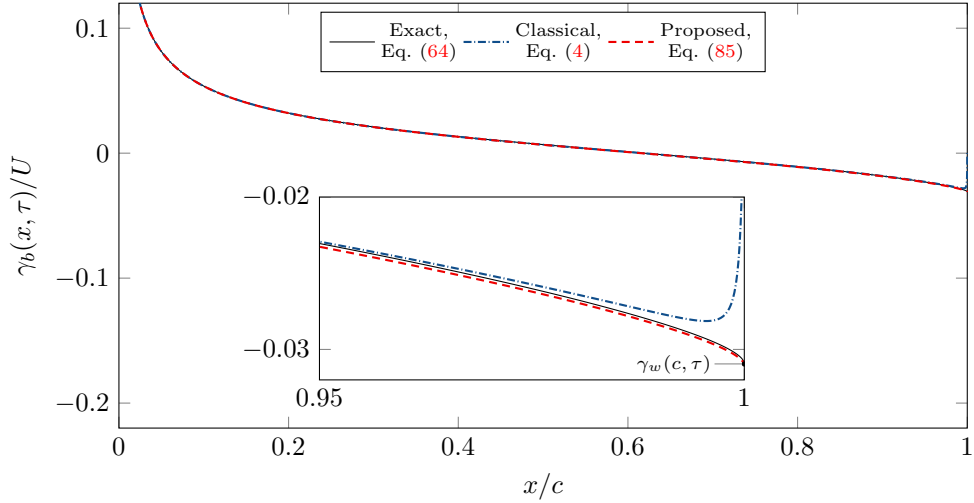

\cref{fig:bound_vorticity_UTAT} compares the bound-vorticity distributions computed through the discrete-vortex method at $\tau=0.5$. The classical formulation again predicts zero bound vorticity at the trailing edge, resulting in a discontinuous vortex-sheet strength. In contrast, the proposed formulation ensures a continuous transition between the bound and wake vorticity distributions, accurately reproducing the analytical solution.

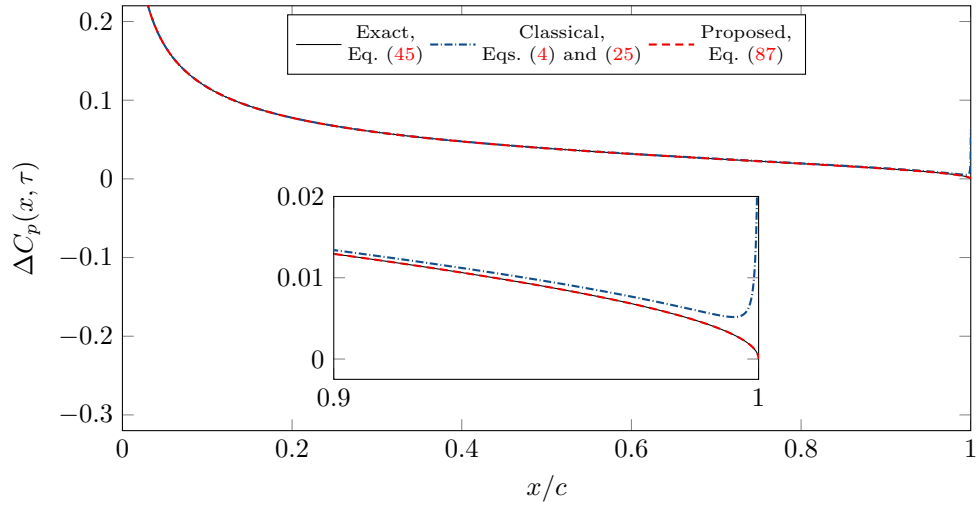
\begin{figure}[!htbp]
\centering
    \newsavebox{\boxCpUTAT}
    \savebox{\boxCpUTAT}
    {\begin{tikzpicture}[>=stealth]
    	\begin{axis}[
    	width=0.45\linewidth,
    	height=0.25\linewidth,
    	xmin=0.9, xmax=1,
        ymin=-0.0025, ymax=0.02,
        xtick = {0.9, 1},
        ytick = {0,0.01,0.02},
        yticklabels={$0$,$0.01$,$0.02$},
        legend style={font=\scriptsize},
        clip = true,
        restrict y to domain = -1:1,
    	]

        \addplot[] table [x = x, y = Cp_Wagner, col sep=comma] {table/Cp_UTAT.txt};

        \addplot[DodgerBlue4,densely dashdotted, thick] table [x = x, y = Cp, col sep=comma] {table/Cp_UTAT.txt};

        \addplot[Red2,densely dashed, thick] table [x = x, y = Cp_mod, col sep=comma] {table/Cp_UTAT.txt};

    	\end{axis}
    \end{tikzpicture}}
    \begin{tikzpicture}[>=stealth]
    	\begin{axis}[
    	width=0.8\linewidth,
    	height=0.45\linewidth,
        xlabel={$x/c$},
    	ylabel={$\Delta C_p(x,\tau)$},
    	xmin=0, xmax=1,
        ymin=-0.32, ymax=0.22,
        xtick = {0,0.2,...,1},
        ytick = {-0.3,-0.2,...,0.2},
        legend columns = 5,
        legend style={font=\scriptsize, at={(0.5,0.98)}, anchor=north, cells={align=center}},
        clip = true,
        restrict y to domain = -1:1,
    	]

        \addplot[] table [x = x, y = Cp_Wagner, col sep=comma] {table/Cp_UTAT.txt};
        \addlegendentry{Exact, \\ \cref{eq:Wagner_Cp}}

        \addplot[DodgerBlue4,densely dashdotted, thick] table [x = x, y = Cp, col sep=comma] {table/Cp_UTAT.txt};
        \addlegendentry{Classical, \\ \cref{eq:glauert_vorticity,eq:pressure_katz2}}

        \addplot[Red2,densely dashed, thick] table [x = x, y = Cp_mod, col sep=comma] {table/Cp_UTAT.txt};
        \addlegendentry{Proposed, \\ \cref{eq:pressure_katz_DVM}}
        
        \draw (axis cs: 0.47,-0.15)         node{\usebox{\boxCpUTAT}};
    	
    	\end{axis}
    \end{tikzpicture}
 \caption{Comparison of the pressure difference distribution computed through the classical and proposed formulations in the discrete-vortex model.}
 \label{fig:Cp_UTAT}
\end{figure}

The influence of the proposed framework on the aerodynamic loading is illustrated in \cref{fig:Cp_UTAT}, where the pressure-difference distributions obtained from the classical and modified formulations are compared. The classical approach predicts a nonzero pressure jump at the trailing edge, violating the zero-loading condition required by the unsteady Kutta condition. Furthermore, this discrepancy becomes increasingly severe at small times due to the contribution of the circulation-rate terms. The proposed formulation regularizes the pressure distribution and ensures that
\begin{equation}
    \Delta C_p(c,\tau)=0,
\end{equation}
thereby recovering the physically expected behavior at the trailing edge.

\begin{figure}[!htbp]
\centering
    \begin{tikzpicture}[>=stealth]
    	\begin{axis}[
    	width=0.8\linewidth,
    	height=0.45\linewidth,
        xlabel={$\tau$},
    	ylabel={$C_l(\tau)$},
    	xmin=0, xmax=2,
        ymin=0.05, ymax=0.08,
        xtick = {0,0.5,...,2},
        ytick = {0.05,0.06,...,0.08},
        yticklabels={$0.05$,$0.06$,$0.07$,$0.08$},
        legend columns = 5,
        legend style={font=\scriptsize, at={(0.5,0.98)}, anchor=north, cells={align=center}},
        clip = true,
        restrict y to domain = 0:3,
    	]

        \addplot[] table [x = t, y = Cl_Wagner, col sep=comma] {table/Cl_UTAT.txt};
        \addlegendentry{Exact, \\ \cref{eq:Wagner_Cl}}

        \addplot[DodgerBlue4,densely dashdotted, thick] table [x = t, y = Cl, col sep=comma] {table/Cl_UTAT.txt};
        \addlegendentry{Classical, \\ \cref{eq:normal_force}}

        \addplot[Red2,densely dashed, thick] table [x = t, y = Cl_mod, col sep=comma] {table/Cl_UTAT.txt};
        \addlegendentry{Proposed, \\ \cref{eq:normal_force_DVM2}}
    	
    	\end{axis}
    \end{tikzpicture}
 \caption{Comparison of the lift coefficient computed through the classical and proposed formulations in the discrete-vortex model.}
 \label{fig:Cl_UTAT}
\end{figure}
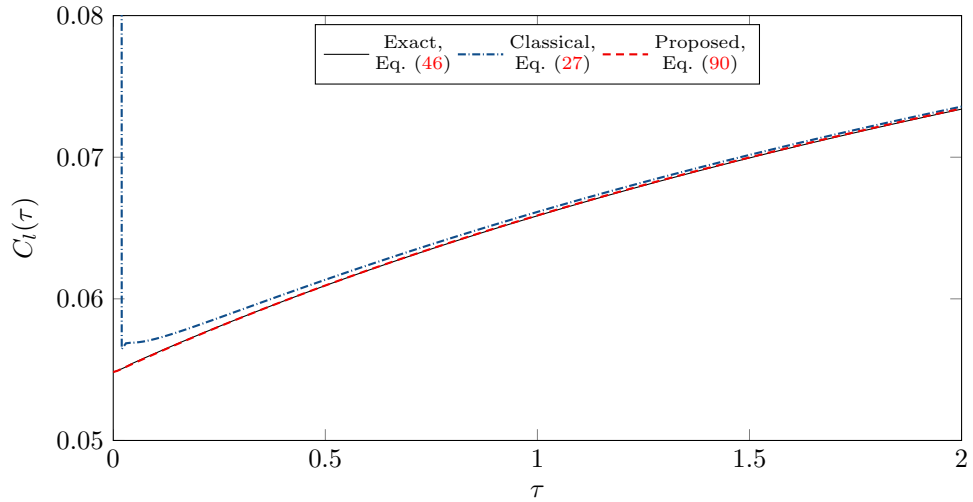

Finally, \cref{fig:Cl_UTAT} presents the lift coefficient predicted by the discrete-vortex method. Although both formulations appear to approach Wagner’s analytical solution at long times, this agreement primarily results from the gradual decay of transient effects. During the early unsteady stage, the classical formulation introduces inconsistencies associated with the trailing-edge pressure loading and the discontinuity of the vortex-sheet strength. In contrast, the proposed formulation remains consistent with the unsteady Kutta condition throughout the transient evolution, ensuring continuous vorticity and zero loading at the trailing edge. Therefore, the principal improvement introduced by the present framework is not merely associated with global force prediction, but rather with the mathematically and physically consistent representation of the unsteady flow in the vicinity of the trailing edge.

\section{Conclusions}

A physically consistent bound-vorticity formulation has been derived for the Wagner problem by enforcing the continuity of the vortex-sheet strength across the trailing edge within the framework of unsteady thin-airfoil theory. The analysis demonstrated that the classical series expansion for the bound vorticity does not converge uniformly at the trailing edge in unsteady flows, leading to discontinuities in the vortex-sheet distribution and inconsistencies in the pressure loading. In particular, the classical formulation imposes a zero bound-vorticity condition at the trailing edge while the shed wake vorticity remains generally nonzero, thereby violating the unsteady Kutta condition.

To address this inconsistency, a recurrence relation for the Wagner coefficients was derived and incorporated into a modified series expansion for the bound vorticity. The resulting formulation yields a uniformly convergent representation of the vortex-sheet strength and recovers the physically expected continuity between the bound and wake vorticity distributions. Furthermore, the pressure difference at the trailing edge is identically zero, ensuring consistency with the zero-loading condition associated with the unsteady Kutta condition.

A discrete-vortex method based on unsteady thin-airfoil theory was developed to investigate the implications of the proposed framework in numerical simulations. The results showed that the modified formulation significantly improves the regularity of the computed vorticity and pressure distributions near the trailing edge. In contrast to the classical expansion, the proposed approach eliminates spurious oscillatory behavior and provides numerically smooth solutions even with relatively few summation terms. The analysis also demonstrated that the apparent agreement of the classical formulation with Wagner’s lift history at large times results primarily from the gradual decay of transient effects rather than from a mathematically consistent treatment of the unsteady flow.

Although the present investigation focuses on the classical Wagner problem, the proposed framework establishes a more robust mathematical foundation for unsteady thin-airfoil theory and discrete-vortex modeling. Since the Wagner solution constitutes the fundamental step-response problem of linear unsteady aerodynamics, the methodology developed here provides a consistent basis for future extensions involving arbitrary kinematics through Duhamel’s principle, as well as more general reduced-order aerodynamic formulations.

\section*{Appendix}

\renewcommand{\thesubsection}{Appendix \Alph{subsection}}
\renewcommand{\theequation}{\Alph{subsection}\arabic{equation}}

\setcounter{equation}{0}

\subsection{Integral in DVM modeling}
\label{apx:theodorsen_model}

The integral of the form
\begin{equation} \label{eq:DVM_integral_0}
    I_n = \int_0^\pi \dfrac{\cos(n\theta)}{\cos\theta + \cosh\vartheta}d\theta
\end{equation}
arises when evaluating the Fourier coefficients $A_n$ in \cref{sec:DVM}. Replacing the modified version of Sears' identity \cite{sears1938systematic},
\begin{equation} \label{eq:sears_identity}
	\dfrac{1}{\cosh\vartheta + \cos\theta} = 2e^{-\vartheta} - 2\sum_{m=2}^\infty (-1)^m e^{-m\vartheta} \dfrac{\sin{(m\theta)}}{\sin\theta},
\end{equation}
into \cref{eq:DVM_integral_0} leads to
\begin{equation} \label{eq:DVM_integral_1}
    I_n =  -\sum_{m=2}^\infty (-1)^m e^{-m\vartheta} \int_0^\pi \left[\dfrac{\sin[(m+n)\theta]}{\sin\theta} + \dfrac{\sin[(m-n)\theta]}{\sin\theta} \right] d\theta.
\end{equation}
The integrals in \cref{eq:DVM_integral_1} can be managed through
\begin{equation}
    \int_0^\pi \dfrac{\sin(p\theta)}{\sin\theta}d\theta = \dfrac{1}{2}\int_0^\pi \left[ \dfrac{\sin(p\theta)\sin\theta}{1 + \cos\theta} + \dfrac{\sin(p\theta)\sin\theta}{1 - \cos\theta} \right]d\theta,
\end{equation}
which is evaluated within Glauert's principal-value integral \eqref{eq:glauert2}:
\begin{equation} \label{eq:DVM_integral_2}
    \int_0^\pi \dfrac{\sin(p\theta)}{\sin\theta}d\theta = \dfrac{\pi}{2} \left[ 1 - (-1)^p\right].
\end{equation}
More generally, \cref{eq:DVM_integral_2} can be written as
\begin{equation} \label{eq:DVM_integral_3}
    \int_0^\pi \dfrac{\sin(p\theta)}{\sin\theta}d\theta =
    \left\{
	\begin{tabular}{rl}
		$\text{sign}(p) \, \pi$, & for $p=\pm 1,\pm 3, \pm 5, \ldots$; \\
        $0$, & otherwise,
	\end{tabular}
	\right.
\end{equation}
which follows that the integral in \eqref{eq:DVM_integral_1} is nonzero for $m = n+2j+1$ ($j=0,1,2,\ldots$):
\begin{equation} \label{eq:DVM_integral_4}
    I_n =  2\pi (-1)^n e^{-(n+1)\vartheta} \sum_{j=0}^\infty e^{-2j\vartheta},
\end{equation}
which can be written as
\begin{equation} \label{eq:DVM_integral_final}
    \int_0^\pi \dfrac{\cos(n\theta)}{\cos\theta + \cosh\vartheta}d\theta = (-1)^n\dfrac{\pi e^{-n\vartheta}}{\sinh\vartheta}, \quad \vartheta \neq 0.
\end{equation}

\section*{Acknowledgments}
The authors acknowledge the financial support of the Brazilian agencies: the National Council for Scientific and Technological Development -- CNPq (grants \#306698/2023-4) and the S\~{a}o Paulo State Research Agency -- FAPESP (grants \#2023/15418-2 and \#2021/09224-5).

\paragraph{CRediT author statement} \textit{George L. S. Torres:} Conceptualization, Methodology, Software, Validation, Writing - Original Draft. \textit{Ashok Gopalarathnam:} Writing - Review \& Editing, Supervision. \textit{Flávio D. Marques:} Conceptualization, Writing - Review \& Editing, Supervision.

\paragraph{Declarations of interest:} None.

\paragraph{Data availability:} Data will be made available from the corresponding author
on request.

\bibliographystyle{unsrtnat}
\bibliography{refs}

\end{document}